\documentclass[aps, prb, reprint, superscriptaddress, preprintnumbers, amsmath, amssymb, letterpaper, floatfix, showpacs]{revtex4-2}
\usepackage{graphicx}
\usepackage{float}
\begin{document}

\preprint{rev. 2.0}

\title{\boldmath Charge-stripe order, antiferromagnetism, and spin dynamics\\ in the cuprate-analog nickelate~La$_4$Ni$_3$O$_8$}

\author{O.~O. Bernal}
\affiliation{Department of Physics and Astronomy, California State University, Los Angeles, California 90032, USA}
\author{D.~E. MacLaughlin}
\affiliation{Department of Physics and Astronomy, University of California, Riverside, California 92521, USA}
\author{G.~D. Morris}
\affiliation{Centre for Molecular and Materials Science, TRIUMF, Vancouver, British Columbia V6T 2A3, Canada}
\author{P.-C. Ho}
\affiliation{Department of Physics, California State University, Fresno, California 93740,USA}
\author{Lei Shu}
\author{C.~Tan}
\author{J.~Zhang}
\author{Z.~Ding}
\author{K.~Huang}
\altaffiliation[Present address: ]{Lawrence Livermore National Laboratory, Livermore, California 94550, USA.}
\affiliation{State Key Laboratory of Surface Physics, Department of Physics, Fudan University, Shanghai 200433, China}
\author{V.~V. Poltavets}
\affiliation{Department of Chemistry and Advanced Materials Research Institute, University of New Orleans, New Orleans, Louisiana 70148, USA}

\date{\today}

\begin{abstract}
We report results of a muon spin rotation ($\mu$SR) study of the cuprate-analog nickelate~La$_4$Ni$_3$O$_8$, which undergoes a transition at 105~K to a low-temperature phase with charge-stripe and antiferromagnetic (AFM) order on square planar NiO$_2$ layers. Zero-field $\mu$SR shows that the AFM transition is abrupt, commensurate and has a quasi-2D character below $\sim$25~K. Comparison of observed muon precession frequencies with Ni dipolar field calculations yields Ni moments~$\lesssim 0.5\mu_B$. Dynamic muon spin relaxation above 105~K suggests critical slowing of Ni spin fluctuations, but is inconsistent with corresponding $^{139}$La NMR results. Critical slowing and an abrupt transition are also observed in the planar cuprate AFM La$_2$CuO$_{4+\delta}$, where they are taken as evidence for weakly interplanar-coupled two-dimensional AFM spin fluctuations, but our $\mu$SR data do not agree quantitatively with theoretical predictions for this scenario when applied to the nickelate.

\end{abstract}


\maketitle

\section{INTRODUCTION} \label{sec:intro}

Bednorz and M\"{u}ller searched for superconductivity in nickel-based systems without success before they discovered superconducting La$_2$BaCuO$_{4+\delta}$~\cite{BNobel}. For many years after, high-$T_c$ superconductivity was found only in cuprate compounds with CuO$_2$-layered structures. It is now well known that besides the cuprates other layered systems (e.g., Fe-pnictide/chalcogenide and BiS$_2$-based materials) also become superconductors, suggesting a potential universality in layered superconductivity. This in turn revives interest in the quest for Cu-analog, Ni-based, superconducting materials. The motivation for studying these materials lies in their potential to help answer current experimental and theoretical questions regarding superconductivity in layered compounds, to assess the universality of layer superconductivity, and to provide clues for where to search for new layered superconductors. 

The trilayer T$^\prime$-type nickelate~La$_4$Ni$_3$O$_8$~\cite{Laco92, PLCM07} is one such material. The crystal structure of this compound involves square planar NiO$_2$ trilayers, isoelectronic to the Cu$^{2+}$O$_2$ layers of the cuprates if the Ni valence were $1+$. Further interest in comparing these systems arises from an antiferromagnetic (AFM) phase transition in La$_4$Ni$_3$O$_8$ at a N\'eel temperature~$T_N = 105$~K into a state with commensurate spin and charge stripes~\cite{PLNC10, ZCPZ16, ZPBZ19u}, similar in some respects to the AFM transition occurring in the cuprates. A charge/spin stripe phase has been observed in the T-type nickelate~La$_{5/3}$Sr$_{1/3}$NiO$_4$~\cite{LeCh97} which, like La$_4$Ni$_3$O$_8$, is 1/3 hole-doped. 

This article reports a muon spin rotation and relaxation ($\mu$SR) study of the magnetic structure and spin dynamics of La$_4$Ni$_3$O$_8$. $\mu$SR is a magnetic resonance technique~\cite{Brew94, *Blun99, *YaDdR11}, conceptually similar to NMR, that uses muon spins implanted into the sample as microscopic probes of their static and dynamic magnetic environment. We find that the 105-K transition is abrupt (discontinuous to within experimental resolution), and that the commensurate AFM configuration is consistent with both magnetic and charge stripe order as found in x-ray~\cite{ZCPZ16} and neutron~\cite{ZPBZ19u} diffraction studies, respectively. Observed spectra of muon spin precession frequencies resemble histograms of calculated Ni dipolar fields at candidate muon sites for specific AFM stripe configurations, with reduced Ni moments 0.4--$0.5\mu_B$. We present $\mu$SR evidence that the magnetic transition at $T_N$ is quasi-2D in nature, in which the trilayers are ordered but fluctuate independently down to $\sim$25~K\@. Below this temperature three-dimensional (3D) static magnetism is observed, with disorder along the normal to the trilayer planes as found by neutron diffraction~\cite{ ZPBZ19u}.

Dynamic muon spin relaxation in the paramagnetic phase suggests critical slowing of Ni spin fluctuations with a divergence at ${\sim}T_N$\@. This is not expected for two-dimensional (2D) Heisenberg magnets, where the critical behavior is due to 2D spin fluctuations that do not order at finite $T$~\cite{CHN88, *CHN89, ChOr90}. In that scenario the 3D transition is due to a weak interplanar coupling, and does not dominate the critical spin dynamics. Evidence for this 2D criticality in the planar cuprate AFM~La$_2$CuO$_{4+\delta}$ is found from NMR studies~\cite{ISYK93, *ISYK94b}. Our $\mu$SR data in La$_4$Ni$_3$O$_8$ do not agree quantitatively with this scenario, and instead suggest the possibility of a second-order transition with a very sharp onset at $T_N$. This is in contrast with the $T = 0$ divergence of the $^{139}$La NMR relaxation rate in La$_4$Ni$_3$O$_8$~\cite{A-WDPG11}, which has been taken as evidence for 2D fluctuations. 

The article is organized as follows: The current understanding of La$_4$Ni$_3$O$_8$ and related systems is reviewed in Sec.~\ref{sec:bkgnd}, with emphasis on the charge/stripe nature of the 105-K transition. Section~\ref{sec:expt} gives experimental details and a brief description of the $\mu$SR technique. Candidate muon stopping sites in the crystal are discussed in Sec.~\ref{sec:musites}, and Sec.~\ref{sec:magorder} reports $\mu$SR results in the ordered phase and their interpretation. Section~\ref{sec:paramag} reports muon spin relaxation in the paramagnetic phase above 105~K\@. Our results are discussed in Sec.~\ref{sec:disc} and summarized in Sec.~\ref{sec:concl}. Appendix~\ref{app:stripes} gives supplementary information on AFM configurations, and Appendix~\ref{app:model} presents a simple model for zero-field $\mu$SR at an abrupt magnetic transition with an inhomogeneous distribution of transition temperatures.

\section{BACKGROUND} \label{sec:bkgnd}

In 1999 Anisimov, Bukhvalov, and Rice~\cite{ABR99} concluded ``Only if the Ni ions are forced into a planar coordination with the O ions can a $S = 1/2$ magnetic insulator be realized with the difficult Ni$^{1+}$ oxidation state and possibly doped with low spin ($S = 0$) Ni$^{2+}$ holes directly analogous to the superconducting cuprates.'' I.e., a system with electronic configuration Ni$^{1+}$/Ni$^{2+}$ would have the same configuration as that of Cu$^{2+}$/Cu$^{3+}$ in the cuprates, but to be a true analog it would also need to be in a square planar coordination with O ions. The Ni$^{1+}$/Ni$^{2+}$ configuration is rare in oxides but it does exist, and conforms with the condition of Anisimov \textit{et~al.}\ in the $T'$-type nickelates~Ln$_{n+1}$Ni$_n$O$_{2n+2}$, Ln = La, Nd, $n =$ 2, 3 and $\infty$~\cite{PLDC06}. In this formula $n$ is the number of NiO$_2$ layers in a multilayer of square-coordinated Ni ions.

Poltavets \textit{et~al.}~\cite{PLNC10} reported a magnetic transition at 105~K for La$_4$Ni$_3$O$_8$ ($n = 3$), and attributed it to Ni-spin inter-trilayer interactions in analogy with undoped La$_2$CuO$_{4+\delta}$. No transition was seen in the magnetization for applied fields lower than 0.1~T, but the observed change in slope of the resistivity $\rho(T)$ and the lambda anomaly in the specific heat at 105~K were taken as evidence for a transition in zero field. At low temperatures $\rho(T)$ exhibits semiconducting behavior~\cite{PLNC10}. $^{139}$La NMR measurements indicated that the transition is to an AFM phase (i.e., 105~K is a N\'eel temperature~$T_N$), further suggesting the possibility of a cuprate analog. 

The results of Ref.~\onlinecite{PLNC10} triggered a good deal of theoretical and experimental work~\cite{A-WDPG11, PaPi10, PaPi12, SDGS-D11, CZGZ12, LZZJ12, Wu13, PAKVS16}. Pardo and Pickett~\cite{PaPi10, *PaPi12} predicted an insulating ground state of the Mott type, and related the transition to binding of $dz^2$ orbitals of Ni ions in outer and middle layers of a trilayer (designated Ni1 and Ni2 respectively). Checkerboard AFM ordering in NiO$_2$ planes was assumed, although powder neutron diffraction has not yielded specific indications for magnetic order~\cite{PLCM07, PLNC10}. A predicted metal-insulator transition under pressure~\cite{PaPi12} was not observed~\cite{CZGZ12}.

Liu \textit{et~al.}~\cite{LZZJ12} and Wu~\cite{Wu13} predicted La$_4$Ni$_3$O$_8$ to be a $C$-type AFM using different theoretical methods. A correlation of the metal/insulator character of the system with dimensionality was found from first-principle calculations in La$_{n+1}$Ni$_n$O$_{2n+2}$ by Liu \textit{et~al.}~\cite{LWJZ14}. The quasi-2D $n = 2$ and 3 varieties were predicted to be molecular insulators in contrast with 3D ($n=\infty$) LaNiO$_2$, which according to their calculated electronic structure is metallic by virtue of Ni-La hybridization. The inclusion of electron correlation effects in first-principles calculations by Patra \textit{et~al.}~\cite{PAKVS16} suggested that the two inequivalent Ni1 and Ni2 atoms in the La$_{4-x}$Sr$_x$Ni$_3$O$_8$ crystal structure are both in a high-spin state, with an average valence of +1.33 independent of $x$ doping.

There were indications, however, that the physics of the nickelates is more complicated than expected. A fit to the temperature dependence of the $^{139}$La NMR relaxation rate $1/T_1$ in La$_4$Ni$_3$O$_8$~\cite{A-WDPG11} required a Korringa term $1/T_1T = \text{const.}$, indicative of a Fermi liquid phase. This is in contrast to the AFM cuprate compounds, which are Mott insulators. In addition, $1/T_1$ in the paramagnetic phase above $T_N$ could be fit as diverging at $T = 0$~\cite{A-WDPG11}, rather than at $T_N$ as expected for a second-order (continuous) transition in 3D\@. A $T = 0$ divergence is characteristic of critical fluctuations of a 2D Heisenberg quantum antiferromagnet~\cite{CHN88, CHN89, ChOr90}, and is observed in the quasi-2D cuprate antiferromagnet~La$_2$CuO$_{4+\delta}$~\cite{ISYK93, ISYK94b}. %In this compound the 3D phase transition at $\sim$320~K is attributed to weak interplanar exchange between large 2D AFM fluctuations (correlation length $\gg$ lattice constant), and is not expected to affect the paramagnetic-phase critical slowing.
%Zero-field Zeeman-split $^{139}$La NQR experiments in La$_2$CuO$_{4+\delta}$~\cite{[{}] [{ and references therein.}] MVBdR94} determined that this  transition is abrupt but continuous, and the magnetic transition in La$_4$Ni$_3$O$_8$ might be similar. But the situation was clearly not well understood. 

More recently, NMR measurements by ApRoberts-Warren \textit{et~al.}~\cite{A-WCDS13} in the paramagnetic phase of the bilayer ($n=2$) $T'$-type compound~La$_3$Ni$_2$O$_6$ revealed very similar behavior of the $^{139}$La relaxation rate to that found previously in trilayer La$_4$Ni$_3$O$_8$. Although they observed no magnetic transition in the bilayer down to 5~K, their results were ascribed to similarity in the electronic structure of the two materials, presumed to be of the 2D variety.

The discussion of the 105-K transition in La$_4$Ni$_3$O$_8$ shifted in 2016, when Zhang, Chen \textit{et~al.}~\cite{ZCPZ16} reported results from synchrotron x-ray diffraction measurements on single crystals. They found a quasi-2D stripe-ordered charge configuration below 105~K, with orientation at 45$^\circ$ to the Ni-O-Ni bonds in NiO$_2$ squares and a 2D supercell~$3\sqrt{2}a \times \sqrt{2}a$, where $a$ is the Ni-Ni nearest-neighbor distance. This is similar to the charge order found in La$_{5/3}$Sr$_{1/3}$NiO$_4$~\cite{LeCh97}, a related 1/3-hole doped single-layer Ruddlesden-Popper nickelate, where the average Ni valence is $2.33+$. The stripe pattern in La$_4$Ni$_3$O$_8$ is two Ni$^{1+}$ ($S = 1/2$) rows followed by one Ni$^{2+}$ ($S = 0$) row, oriented at 45$^\circ$ to the planar Ni-O bonds, with staggered stacking along the $c$-axis. The 2:1 ratio preserves the formal Ni valence $1.33+$. All three layers in a trilayer possess stripes. The magnetic structure was not obtained in this study, which sampled only charge degrees of freedom.

\begin{figure} 
\includegraphics[clip=,width=0.35\textwidth]{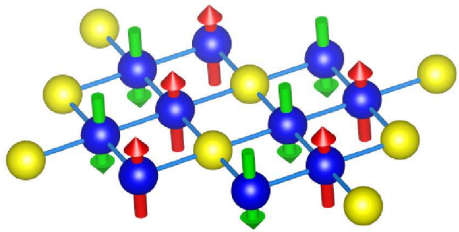}
\caption{\label{fig:stripes} Charge stripe~\cite{ZCPZ16} and proposed AFM moment configurations~\cite{BPPN16} on a nickel layer in La$_4$Ni$_3$O$_8$. Blue spheres: Ni$^{1+}$ ions ($S = 1/2)$. Yellow spheres: Ni$^{2+}$ ions ($S = 0$). Arrows: AFM ordered moments. O$^{2-}$ ions are not shown.}
\end{figure}

\textit{Ab~initio} calculations by Botana \textit{et~al.}~\cite{BPPN16} yielded stripe charge order from a combination of structural distortion and AFM order, in which Ni$^{2+}$ ions have spin $S = 0$, and Ni$^{1+}$ $S = 1/2$ ions order antiferromagnetically, similar to the spin-1/2 insulating antiferromagnetism of the cuprate parent materials. In this result the valences of corresponding Ni1 and Ni2 ions in a trilayer are the same. Two 2D supercells of the observed stripe pattern and the proposed 2D AFM order in a single Ni layer~\cite{BPPN16} are shown in Fig.~\ref{fig:stripes}. 

Very recently Zhang, Pajerowski \textit{et~al.}~\cite{ZPBZ19u} reported results of neutron Bragg diffraction experiments on La$_4$Ni$_3$O$_8$. Their measurements confirm the magnetic ground state, with stripe AFM order commensurate with the charge stripes in individual trilayers but little or no correlation between trilayers. In contrast to La$_{5/3}$Sr$_{1/3}$NiO$_4$ and to most other transition metal oxides, the charge and spin stripes form simultaneously at 105~K.

\section{EXPERIMENT} \label{sec:expt}

\subsection{Sample} \label{sec:sample} 

Polycrystalline La$_4$Ni$_3$O$_8$ was prepared as described previously~\cite{PLCM07} and characterized by powder x-ray diffraction, x-ray absorption spectroscopy, magnetization, specific heat, resistivity, and NMR measurements~\cite{PLCM07, PLNC10, A-WDPG11}. For our $\mu$SR experiments we used enough material to fully cover a circular area of about 1~cm in diameter and 1~mm thickness.

\begin{figure}
\includegraphics[clip=,width=0.40\textwidth]{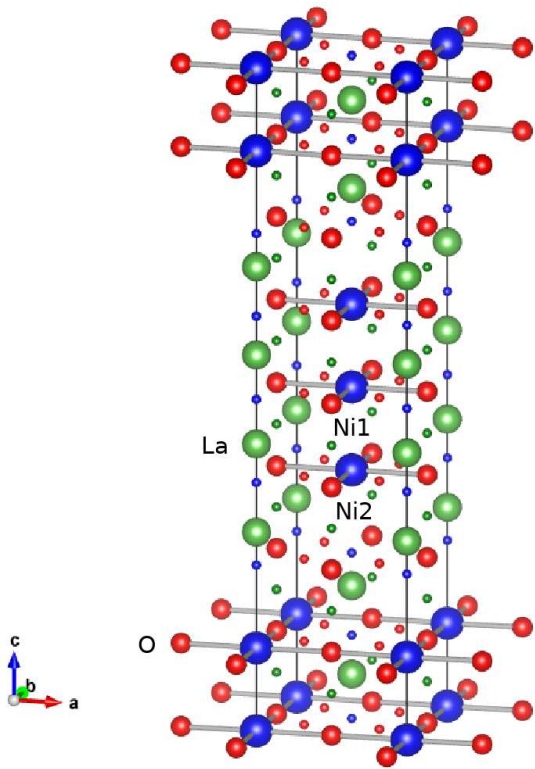}
\caption{\label{fig:LNO} Unit cell of La$_4$Ni$_3$O$_8$ (space group~$I4/mmm$, no.~139). Blue spheres: Ni. Red spheres: O. Green spheres: La. Small spheres: candidate muon sites (cf.\ Sec.~\ref{sec:musites}, Table~\ref{tab:musites}). Red: $\mu$1--$\mu$3. Green: $\mu$4--$\mu$5. Blue: $\mu$6--$\mu$8.}
\end{figure}

The two-trilayer tetragonal unit cell of La$_4$Ni$_3$O$_8$ (space group~$I4/mmm$, no.~139) is shown in Fig.~\ref{fig:LNO}. 
There are two inequivalent Ni sites (Ni1 and Ni2) and three inequivalent oxygen sites, two of which coordinate the Ni ions in the planar NiO$_2$ layers. The third oxygen sites form separate square-coordinated layers, with La ions above and below the square centers forming flourite La-O blocks that separate and isolate one NiO$_2$ trilayer from the next. NiO$_2$ layers within a trilayer are separated by La-only planes. The candidate muon sites shown in Fig.~\ref{fig:LNO} are discussed in Sec.~\ref{sec:musites}.

\subsection{\boldmath $\mu$SR experiments} 

In the $\mu$SR technique~\cite{Brew94, Blun99, YaDdR11}, 100\% spin-polarized muons are implanted into a sample, where they decay [$\mu^\pm \to e^\pm + \nu_e(\overline{\nu}_e) + \overline{\nu}_\mu(\nu_\mu)$] with a mean lifetime 2.197~$\mu$s. Usually positive muons ($\mu^+$) are used, which stop in interstitial sites~\footnote{Negative muons go into tight Bohr orbits around host nuclei and are less sensitive to their magnetic environment than interstitial positive muons.}. The decay positron is preferentially emitted in the direction of the $\mu^+$ spin at the time of decay. Thus the time-dependent positron count-rate asymmetry~$A(t)$ (the \emph{asymmetry spectrum}), where $t$ is the time after implantation, directly yields the time evolution of the ensemble $\mu^+$ spin polarization~$G(t)$ [$G(0) \equiv 1$]: $A(t) = A_0G(t)$, where the initial asymmetry~$A_0 = 0.2$--0.3 is instrument-dependent. Information on local magnetic behavior is contained in $G(t)$. Measurement of a single asymmetry spectrum typically requires $\gtrsim 10^7$ events.

$\mu^+$SR experiments on La$_4$Ni$_3$O$_8$ were carried out with the LAMPF spectrometer at the M20D muon beam line, TRIUMF, Vancouver, Canada, over the temperature range~1.2--300~K, in zero magnetic field (ZF) and in longitudinal applied fields (LF) (field parallel to the initial $\mu^+$ spin polarization) up to 4~kOe. Data were analyzed using the Paul Scherrer Institute \texttt{musrfit} fitting program~\cite{SuWo12} and the TRIUMF \textsc{physica} programming environment~\footnote{\texttt{http://computing.triumf.ca/legacy/physica/}}.

\section{\boldmath $\mu^+$ STOPPING SITES}\label{sec:musites}

$\mu^+$ precession frequencies in ordered magnets yield static magnetic field values at the $\mu^+$ stopping sites, which can help to characterize the magnetic order if these sites are known. Implanted positive muons in oxides are likely to stop near O$^{2-}$ ions, and in La$_4$Ni$_3$O$_8$ with stripe order the large unit cell yields a considerable number of such candidate sites. To date there are no reported calculations of site locations or relative occupations in La$_4$Ni$_3$O$_8$ using, for example, DFT theory~\cite{[{For a review, see }] BoDR16}. 

In lieu of such information, we consider candidate $\mu^+$ sites that are symmetric with respect to oxygen near neighbors. Eight such sites that are crystallographically inequivalent are shown in Fig.~\ref{fig:LNO} and listed in Table~\ref{tab:musites}.
\begin{table*}
\caption{\label{tab:musites} Candidate symmetric $\mu^+$ stopping sites in La$_4$Ni$_3$O$_8$. Coordinates are fractions of lattice parameters in the $I4/mmm$ and $F4/mmm$ settings (see text). Coordinates and O$^{2-}$ distances from lattice data of Ref.~\onlinecite{PLCM07}.}
\begin{ruledtabular}
\begin{tabular}{lrcrcrrcc}
 & \multicolumn{3}{c}{$I4/mmm$} & Wycoff & \multicolumn{2}{c}{$F4/mmm$} & O$^{2-}$ distance & Location \\[-3pt]
$\mu^+$ site & $x$ & $y$ & $z$\footnotemark[1] & designation & $x$ & $y$ & (\AA) \\
\colrule
$\mu 1$ & $\frac{1}{4}$ & $\frac{1}{4}$ & 0 & $8h$ & 0 & $\frac{1}{4}$ & 1.4039 & nn O in $ab$ plane \\
$\mu 2$ & $\frac{1}{4}$ & $\frac{1}{4}$ & 0.1256 & $16m$ & 0 & $\frac{1}{4}$ & $''$ $''$ & $''$ \qquad $''$ \\
$\mu 3$ & $\frac{1}{4}$ & $\frac{1}{4}$ & $\frac{1}{4}$ & $8f$ & 0 & $\frac{1}{4}$ & $''$ $''$ & $''$ \qquad $''$ \\
$\mu 4$ & 0 & $\frac{1}{2}$ & 0.1878 & $8g$ & $\frac{1}{4}$ & $\frac{1}{4}$ & 1.6238 & nn O along $c$ axis \\
$\mu 5$ & 0 & $\frac{1}{2}$ & 0.0628 & $8g$ & $\frac{1}{4}$ & $\frac{1}{4}$ & 1.6394 & $''$ \qquad $''$ \\
$\mu 6$ & $\frac{1}{2}$ & $\frac{1}{2}$ & 0 & $2b$ & 0 & $\frac{1}{2}$ & 1.9854 & centered in O square \\
$\mu 7$ & $\frac{1}{2}$ & $\frac{1}{2}$ & 0.1256 & $4e$ & 0 & $\frac{1}{2}$ & $''$ $''$ & $''$ \qquad $''$ \\
$\mu 8$ & $\frac{1}{2}$ & $\frac{1}{2}$ & $\frac{1}{4}$ & $4e$ & 0 & $\frac{1}{2}$ & $''$ $''$ & $''$ \qquad $''$ \\
\end{tabular}
\end{ruledtabular}
\end{table*}
It is customary~\cite{SDGS-D11, ZCPZ16, ZPBZ19u, BPPN16} to describe magnetic and stripe structure in La$_4$Ni$_3$O$_8$ using the setting~$F4/mmm$, in which basal-plane unit vectors are rotated 45$^\circ$ to define a face-centered square lattice of Ni ions. The lattice constants in these directions are then $\sqrt{2}$ times the Ni-Ni nearest neighbor distance~$a$ in the basal plane. The lattice constant~$c$ remains the same as in the $I4/mmm$ setting, and the unit cell volume is doubled. The observed charge-stripe supercell~\cite{ZCPZ16} is $3\sqrt{2}a \times \sqrt{2}a \times c$, where the stripe direction is the $b$ axis (cf.\ Fig.~\ref{fig:stripes}). Candidate $\mu^+$ sites are given in Table~\ref{tab:musites} for both settings. 
 
\begin{figure}
\includegraphics[clip=,width=0.40\textwidth]{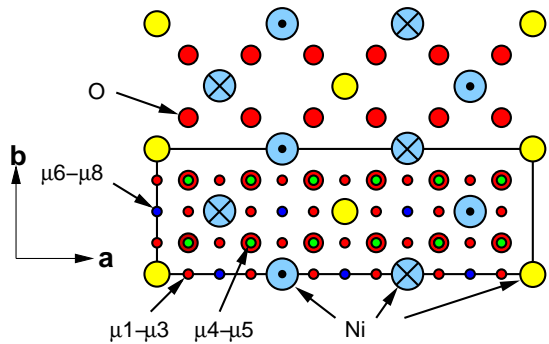}
\caption{\label{fig:musites} 2D AFM structure of Fig.~\ref{fig:stripes}~\cite{BPPN16} and projections of candidate $\mu^+$ stopping sites in La$_4$Ni$_3$O$_8$ (Fig.~\ref{fig:LNO} and Table~\ref{tab:musites}) onto the $ab$ plane in the $F4/mmm$ setting of the $3\sqrt{2}a \times \sqrt{2}a$ 2D stripe supercell. Large circles: Blue: Ni$^{1+}$ ($S = 1/2$) ($\odot =$ up-spins, $\otimes =$ down-spins). Yellow: Ni$^{2+}$ ($S = 0$). Red: O$^{2-}$. Small circles: $\mu^+$ sites from Table~\ref{tab:musites}. Rectangle: supercell boundary.}
\end{figure}

In the $I4/mmm$ tetragonal unit cell there are a total of 58 such sites (the sum of the Wycoff multiplicities in Table~\ref{tab:musites}). With two trilayers per $I4/mmm$ unit cell, there are 58 sites per trilayer in the $F4/mmm$ setting and $3 \times 58 = 174$ sites per trilayer in a stripe supercell. The two trilayers are crystallographically equivalent, so that it is sufficient to consider $\mu^+$ sites for a single trilayer provided this equivalence extends to the ordered magnetic structure. Figure~\ref{fig:musites} shows two $F4/mmm$ supercells of La$_4$Ni$_3$O$_8$ viewed along the $c$ axis, with the stripe AFM configuration of Fig.~\ref{fig:stripes}~\cite{BPPN16} and $\mu^+$ sites from Table~\ref{tab:musites}. 
The number of magnetically inequivalent sites in a full 3D AFM configuration is clearly greater in stripe models than for AFM order without stripes. 

The $\mu^+$ sites of Table~\ref{tab:musites} may not be realistic: they would be unstable if the O$^{2-}$ ions were classical point charges, and our calculations do not take account of off-site ionic positions, intrinsic to the La$_4$Ni$_3$O$_8$ crystal structure~\cite{BPPN16} or due to the $\mu^+$ charge. The candidate sites sample a large fraction of the unit cell, however (Fig.~\ref{fig:LNO}), so that calculated dipolar fields at these sites in the magnetically ordered phase can be taken as rough guides to the local field distribution.

\section{ANTIFERROMAGNETIC CONFIGURATIONS} \label{sec:magorder}

\subsection{Experimental results} \label{sec:expres}

\begin{figure}
\includegraphics[clip=,width=0.40\textwidth]{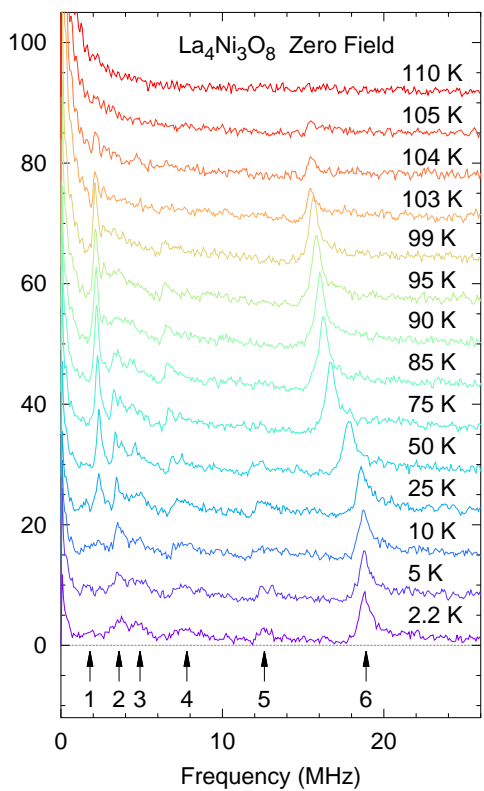}
\caption{\label{fig:res-FFT} Fourier-transform spectra from ZF-$\mu$SR asymmetry data (cf.\ Fig.~\ref{fig:exp-magord-asy}) in La$_4$Ni$_3$O$_8$, $2.2~\text{K} \leqslant T \leqslant 110$~K\@. The spectra are offset for clarity. Arrows: discrete frequencies at 2.2~K.}
\end{figure}

Figure~\ref{fig:res-FFT} shows Fourier transforms of ZF-$\mu$SR asymmetry data from La$_4$Ni$_3$O$_8$ from 2.2~K to 110~K\@. 
There are at least six discrete peaks with nonzero frequency, corresponding to oscillations in the asymmetry spectra due to $\mu^+$ spin precession in well-defined magnetic fields. The peaks are clearly visible at 50~K; we label them from 1 to 6 on the abscissa of Fig.~\ref{fig:res-FFT}. In the temperature range 99--105~K the frequency of Peak~6 ($\sim$19~MHz) hardly changes, but its amplitude decreases smoothly to nearly zero. Several of the peaks are broad, and probably contain unresolved frequencies from multiple sites. 

\begin{figure}
\includegraphics[clip=,width=0.40\textwidth]{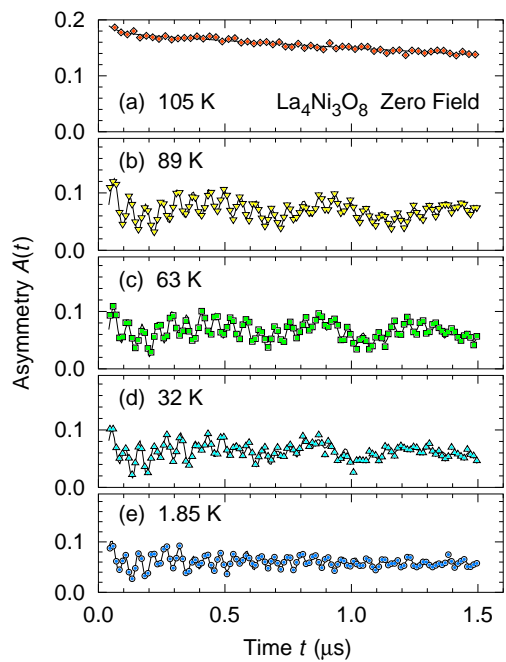}
\caption{\label{fig:exp-magord-asy} Representative ZF-$\mu$SR asymmetry spectra in La$_4$Ni$_3$O$_8$, $T \leqslant T_N = 105$~K\@. Curves: fits of Eq.~(\ref{eq:multifrq}) to the data.}
\end{figure}

A representative series of ZF-$\mu$SR asymmetry spectra for temperatures between 1.85~K and 105~K (the AFM phase) is shown in Fig.~\ref{fig:exp-magord-asy}.
The curves in Fig.~\ref{fig:exp-magord-asy} are fits to the data using the exponentially-damped multi-frequency function
\begin{equation} \label{eq:multifrq}
G(t) = f_0\exp(-\lambda_0t) + \sum_{i=1}^6 f_i\exp(-\lambda_it)\cos(\omega_it+\phi_i)
\end{equation} 
for the $\mu^+$ spin polarization~\footnote{For ZF measurements in a random powder with static local fields, 1/3 of the initial $\mu^+$ polarization relaxes with zero frequency~\protect\cite{KuTo67, *HUIN79}. A $\omega = 0$ component is also present in an antiferromagnet if a muon site is symmetric with respect to the sublattices, since this results in cancellation of local fields.}. The frequencies~$\omega_i$ give the local magnetic fields~$\omega_i/\gamma_\mu$, where $\gamma_\mu= 8.5156 \times 10^4~\mathrm{s}^{-1}~\mathrm{G}^{-1}$ is the gyromagnetic ratio of the muon. The rates~$\lambda_i$ characterize line broadening, dominated by inhomogeneity in these fields. The amplitudes~$f_i$ are related to $\mu^+$ site occupation probabilities, and are not expected to depend on temperature. The polycrystalline nature of our sample is not involved in the ZF results shown in Fig.~\ref{fig:exp-magord-asy}, except that the $f_i$ are averaged over local-field orientations relative to the initial $\mu^+$ polarization direction.

\begin{figure}
\includegraphics[clip=,width=0.40\textwidth]{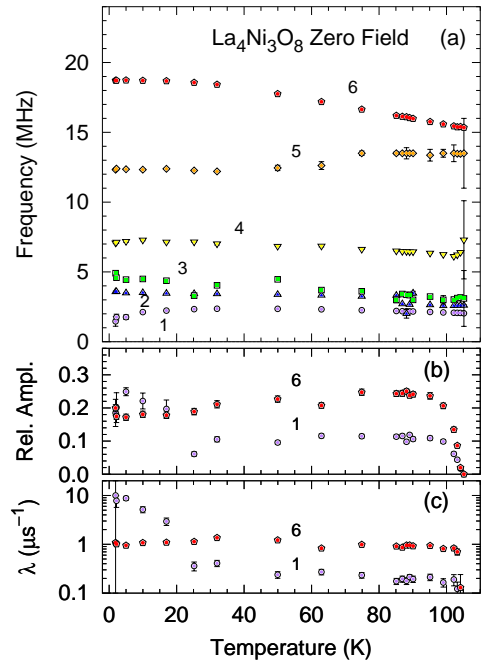}
\caption{\label{fig:res-6freq} Temperature dependence of parameters in La$_4$Ni$_3$O$_8$ from fits of Eq.~(\ref{eq:multifrq}) to ZF-$\mu$SR asymmetry data (cf.\ Fig.~\ref{fig:exp-magord-asy}). Numbers indicate peaks in Fig.~\ref{fig:res-FFT}. (a)~$\mu^+$ precession frequencies. (b)~Relative oscillation amplitudes of Peaks~1 and 6 (Fig.~\ref{fig:res-FFT}). (c) Exponential damping rates of Peaks~1 and 6.}
\end{figure}

Frequencies obtained from fits of Eq.~(\ref{eq:multifrq}) to asymmetry data are plotted vs temperature in Fig.~\ref{fig:res-6freq}~(a) and similarly labeled from 1 to 6. 
The rather smooth differences in temperature dependence of the frequencies between $T_N$ and $\sim$25~K % [Fig.~\ref{fig:res-6freq}~(a)] 
is evidence for a gradual Ni spin reorientation below the transition.

The amplitudes and rates of Peak~1 ($\sim$2~MHz) and Peak~6 are compared in Figs.~\ref{fig:res-6freq}(b) and \ref{fig:res-6freq}(c). They behave quite differently: below $\sim$90~K the amplitude and rate of Peak~6 are nearly constant, whereas for Peak~1 they both increase substantially with decreasing temperature below $\sim$25~K\@. At lower temperatures Peak~1 is hard to see in the Fourier spectra (Fig.~\ref{fig:res-FFT}), because its increase in time-domain (asymmetry) amplitude by a factor of $\sim$2 [Fig.~\ref{fig:res-6freq}(b)] (probably an artifact due to overlapping adjacent peaks) is offset~\footnote{Due to the general Fourier-transform proportionality between time-domain amplitudes and frequency-domain areas, increased broadening causes a loss of frequency-domain amplitude.} by a damping-rate (linewidth) increase by a factor of $\sim$30 [Fig.~\ref{fig:res-6freq}(c)]. This behavior is discussed in detail in Sec.~\ref{sec:spinconf}.

\subsection{\boldmath Models of AFM order: dipolar-field calculations} \label{sec:modelcalc}

This section describes lattice-sum calculations of Ni-moment dipolar magnetic fields at the candidate $\mu^+$ stopping sites of Table~\ref{tab:musites} for a number of candidate AFM configurations, with and without charge stripes. The calculations assume the same moment on all moment-bearing Ni$^{1+}$ ions in a trilayer, consistent with the neutron diffraction results~\cite{ZPBZ19u}. These dipolar fields are summed over a sphere of radius 200~\AA\ centered on the $\mu^+$ site, after determining that increasing this radius does not affect the results appreciably. The Ni-ion moment is scaled for rough agreement of the dipolar-field frequencies with a typical experimental spectrum ($T = 50$~K, cf.\ Fig.~\ref{fig:res-FFT}). 

\subsubsection{AFM configurations with charge stripes} \label{sec:modelcalcstripes}

\begin{figure}
\includegraphics[clip=,width=0.40\textwidth]{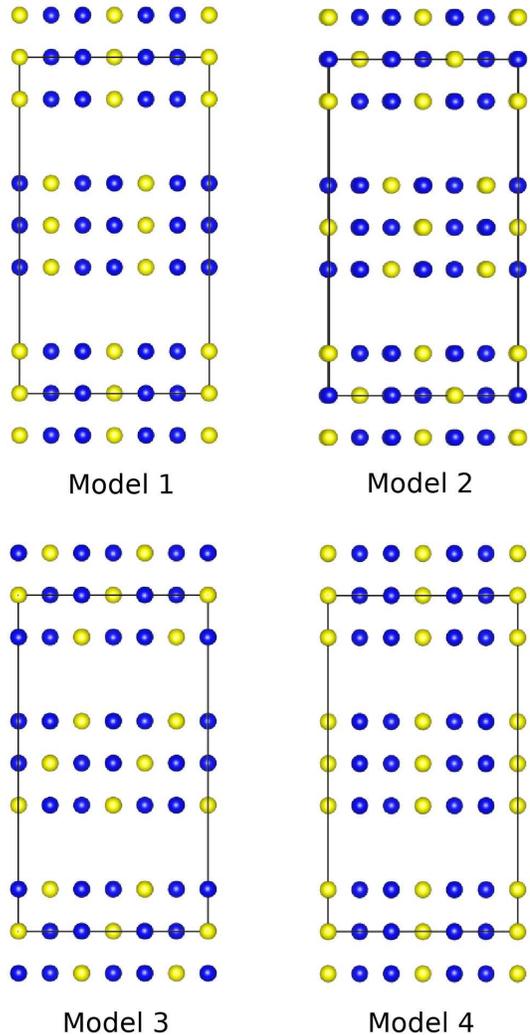}
\caption{\label{fig:models} Models for charge stripes in La$_4$Ni$_3$O$_8$. Numbering as in Ref.~\onlinecite{ZCPZ16}. Blue spheres: Ni$^{1+}$ ions ($S=1/2$). Yellow spheres: Ni$^{2+}$ ions ($S = 0$). Rectangles: $3\sqrt{2}a \times \sqrt{2}a \times c$ supercell outlines. The view is along the stripe ($b$) direction.}
\end{figure}

\paragraph{Stripe models.} For completeness, we consider the four charge stripe models discussed by Zhang, Chen \textit{et~al.}~\cite{ZCPZ16}, which differ in the positions of the stripes on the individual layers. These models are shown in Fig.~\ref{fig:models}, where the view is along the stripe direction (the $b$ direction in the $F4/mmm$ setting), and a supercell boundary is shown. 
The model numbering (1--4) of Ref.~\onlinecite{ZCPZ16} is used in the following. The stripes in a trilayer are stacked along the $c$ axis in Models 1 and 4 and staggered in Models 2 and 3. The Coulomb repulsion energy would seem to be lower for the latter, but x-ray diffraction data favor the stripe structure of Model~1~\cite{ZCPZ16}. The latter is also used for AFM configurations in the neutron diffraction study of Ref.~\onlinecite{ZPBZ19u}. Model~4 is assumed in the \textit{ab~initio} treatment of charge and AFM order in La$_4$Ni$_3$O$_8$ of Botana \textit{et~al.}~\cite{BPPN16}, where it is argued that the weak coupling between trilayers does not play a significant role in the electronic structure. 

The calculations and neutron diffraction results cited above~\cite{BPPN16, ZPBZ19u} both yield the 2D AFM structure shown in Fig.~\ref{fig:stripes}, with Ni moments aligned along the $c$ axis. We have calculated dipolar fields at $\mu^+$ sites for 16 3D AFM structures, obtained by placing 2D AFM layers with this structure on each of the four stripe models in four configurations, designated A--D in the following. In Configuration~A the moment orientations are those of Fig.~\ref{fig:stripes} on all layers. In Configuration~B all moments on alternate trilayers of Configuration~A are inverted. In Configuration~C the Ni2 (middle layer) moments of Configuration~A are inverted for all layers, yielding intra-trilayer AFM ordering. Configuration~D can be obtained in either of two ways: by inverting the Ni2 moments of Configuration~B, or by inverting alternate trilayers of Configuration C\@.

\begin{figure} [ht]
\includegraphics[clip=,width=0.40\textwidth]{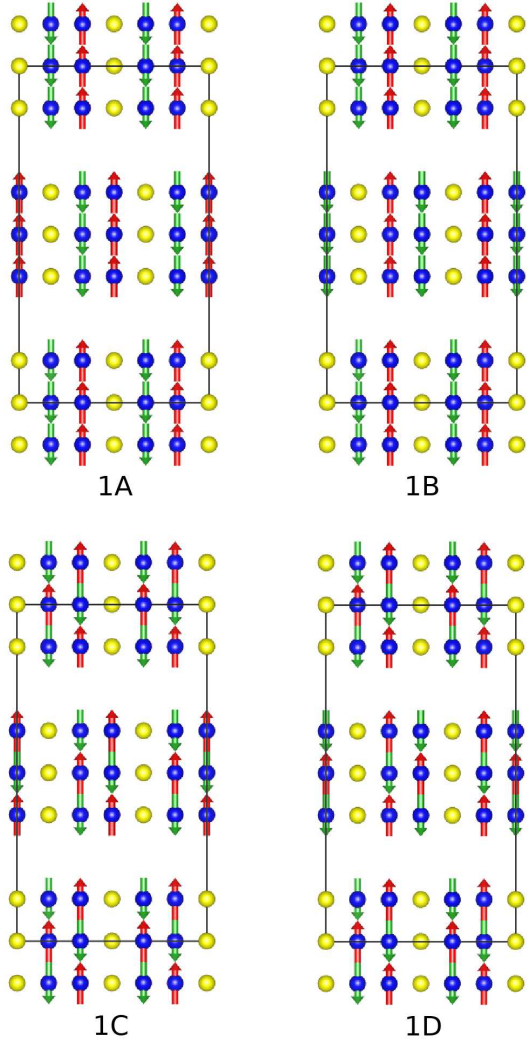}
\caption{\label{fig:AFM-mod1} AFM moment Configurations A--D in La$_4$Ni$_3$O$_8$ on charge stripe Model~1~\cite{ZCPZ16}. Blue spheres: Ni$^{1+}$ ions ($S=1/2$). Yellow spheres: Ni$^{2+}$ ions ($S = 0$) Rectangles: $3\sqrt{2}a \times \sqrt{2}a \times c$ supercell outlines. Arrows: AFM ordered moments. The view is along the stripe ($b$) direction.}
\end{figure}
\begin{figure} [ht]
\includegraphics[clip=,width=0.40\textwidth]{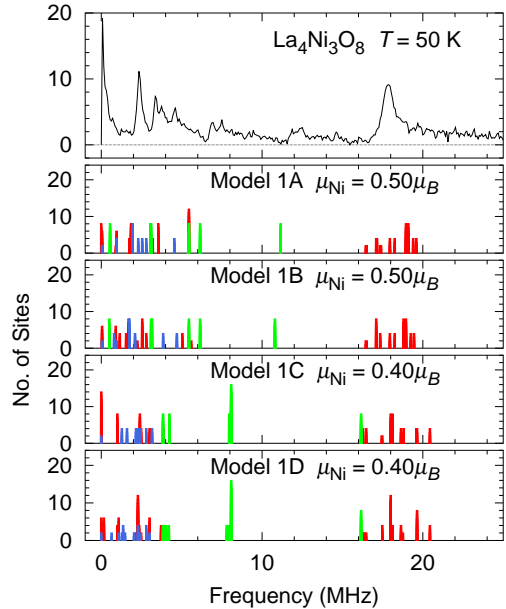}
\caption{\label{fig:disc-spct1} Histograms of dipolar-field $\mu^+$ frequencies from AFM stripe Models~1A--1D compared to 50-K $\mu$SR FT spectrum in La$_4$Ni$_3$O$_8$, $T = 50$~K\@. Colors indicate $\mu^+$ site groups (Table~\ref{tab:musites}): Red: $\mu 1$--$\mu 3$. Green: $\mu 4$--$\mu 5$. Blue: $\mu 6$--$\mu 8$.}
\end{figure}
We discuss Model~1 below. Models~2--4 are described in Appendix~\ref{app:stripes}. 

\paragraph{Dipolar fields in stripe Model~1.} \label{sec:mod1fields} The four AFM moment configurations are shown in Fig.~\ref{fig:AFM-mod1} for charge stripe Model~1, and histograms of frequencies from $\mu^+$-site dipolar-field calculations for these configurations are compared with the observed 50-K spectrum in Fig.~\ref{fig:disc-spct1}. 
The calculated histograms assume equal stopping probabilities for all $\mu^+$ sites, which may not be the case, and do not take into account possible transferred $\mu^+$ hyperfine fields. A rough correspondence between the calculated histograms and observed spectra is obtained with Ni$^{1+}$ moments 0.4--0.5$\mu_B$, somewhat smaller than the predicted values 0.6--0.7$\mu_B$~\cite{BPPN16}. A high-frequency group of peaks is separated from a low-frequency group by a region with less spectral weight. The comparison suggests that the high-frequency peak is formed from a group of discrete frequencies. These are mainly from $\mu1$ sites, which is not surprising as they are the closest to Ni ions (Fig.~\ref{fig:musites}).

There are a number of general features of the calculated histograms (Fig.~\ref{fig:disc-spct1} and Appendix~\ref{app:stripes}). Histograms for stripe Models~1 and 4 are similar, as are those for stripe Models~2 and 3, due to stacked stripes in the former compared to offset stripes in the latter. For a given model, only small differences and no change of Ni moment are found between A and B histograms, and also between C and D histograms. Inverting alternate trilayers changes the $\mu^+$-site fields considerably less than inverting Ni2 moments within each trilayer. 

Reference~\onlinecite{ZPBZ19u} concludes that trilayers are essentially uncorrelated along the $c$ axis, corresponding to a random mix of Configurations 1A and 1B (or 1C and 1D)\@. From Fig.~\ref{fig:disc-spct1}, such a random mix would have little effect on the higher frequencies, which arise from $\mu^+$ sites in or near the trilayers. There are visible changes of the lower frequencies associated with sites far from trilayers. This behavior may have been observed, as described below in Sec.~\ref{sec:spinconf}.

For all stripe models a calculated peak or group is found near the observed small peak at $\sim$12~MHz for AFM Configurations~A and B but not for Configurations~C and D; agreement with the observed spectrum is better for the former configurations. This is not strong evidence for them, however, due to the approximate nature of the treatment and the uncertainty in the $\mu^+$ sites. Nevertheless, we note the discrepancy between this result and the neutron scattering data~\cite{ZPBZ19u}, which favor the AFM intra-trilayer order of Configurations~1C and 1D (trilayer model 5 of Ref.~\onlinecite{ZPBZ19u}, Supplementary Information) over the FM intra-trilayer order of Configurations 1A and 1B (their model 6).

\subsubsection{AFM order without stripes?} \label{sec:nostripes}
\begin{figure} [ht]
\includegraphics[clip=,width=0.40\textwidth]{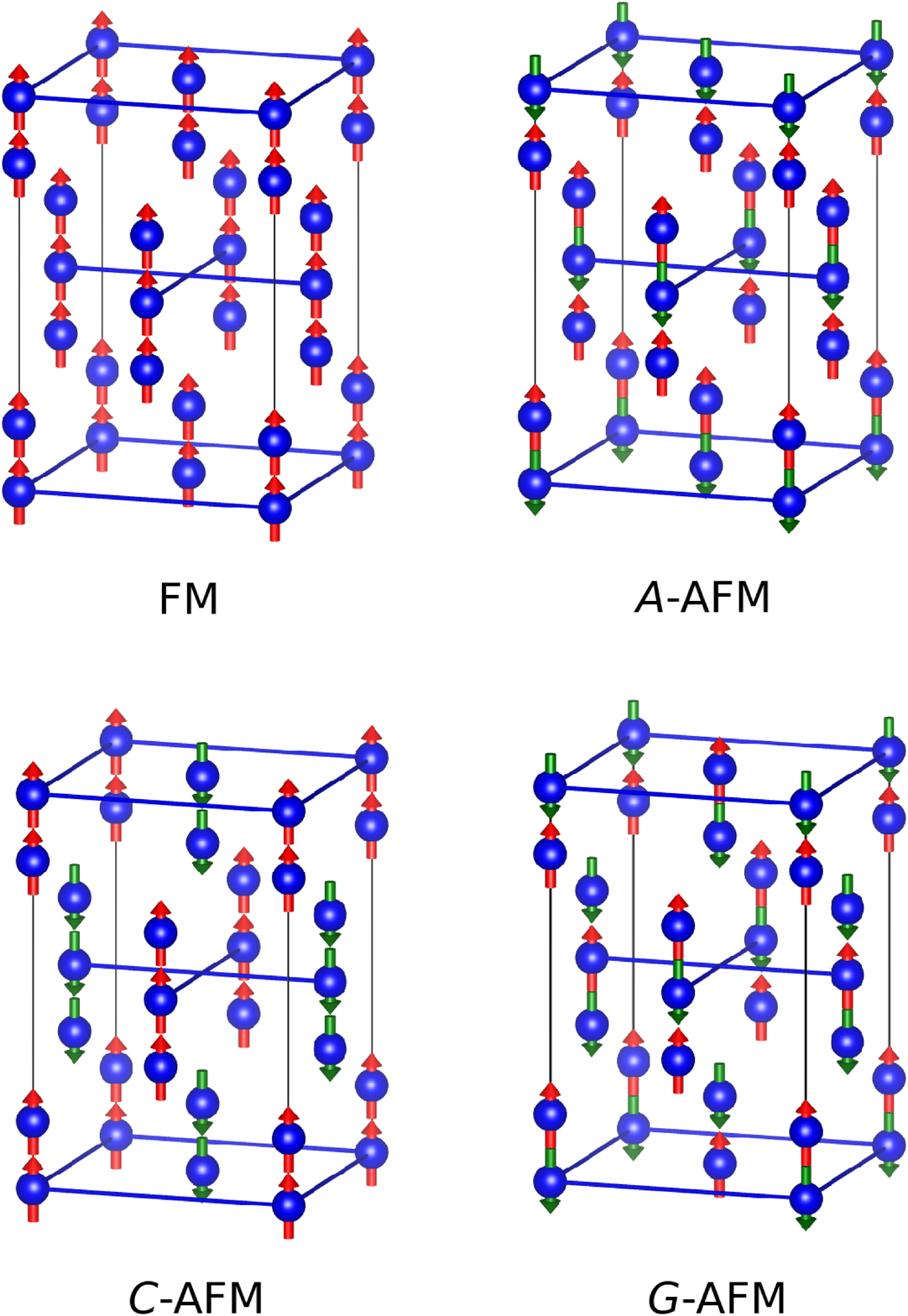}
\caption{\label{fig:homo-0} AFM Ni-ion moment configurations without stripes in La$_4$Ni$_3$O$_8$, $F4/mmm$ setting. $A$-AFM: FM intra-trilayer, AFM inter-trilayer. $C$-AFM: AFM intra-trilayer, FM inter-trilayer. $G$-AFM: AFM intra-trilayer, AFM inter-trilayer.}
\end{figure}
\begin{figure} [ht]
\includegraphics[clip=,width=0.40\textwidth]{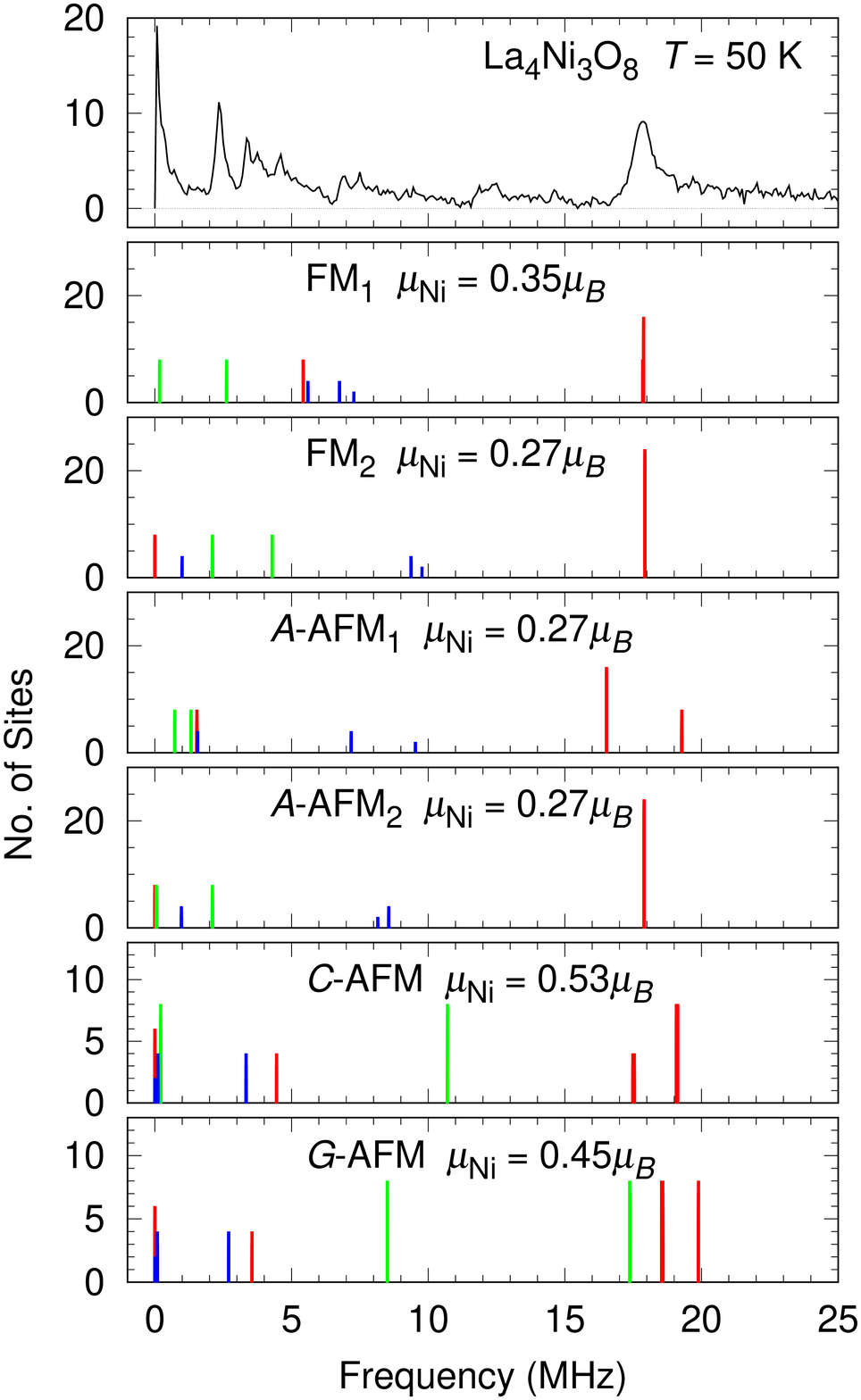}
\caption{\label{fig:homo-spct} Histograms of dipolar-field $\mu^+$ frequencies from AFM configurations without stripes (Fig.~\ref{fig:homo-0}) compared to 50-K $\mu$SR FT spectrum in La$_4$Ni$_3$O$_8$. See text for notation. Colors as in Fig.~\ref{fig:disc-spct1}.}
\end{figure}
For completeness we also consider AFM configurations without stripes (although these are ruled out by the neutron diffraction results~\cite{ZPBZ19u}). Using the $F4/mmm$ setting, Fig.~\ref{fig:homo-0} shows four such configurations in La$_4$Ni$_3$O$_8$: the ferromagnetic (FM) configuration (for completeness), and three AFM configurations~\cite{SDGS-D11, LZZJ12, Wu13, PAKVS16}. Figure~\ref{fig:homo-spct} compares frequency histograms from dipolar field calculations for the configurations of Fig.~\ref{fig:homo-0} with the observed 50-K spectrum. The subscripts 1 and 2 for the FM and $A$-AFM configurations designate alternate trilayer moments as pictured in Fig.~\ref{fig:homo-0} or inverted, respectively. Alternate-trilayer inversion followed by 90$^\circ$ rotation about the $c$ axis is a symmetry operation for the $C$-AFM and $G$-AFM configurations, and therefore does not affect the dipolar-field histogram. 

These histograms have a rough correspondence with those for the stripe-model AFM configurations of the previous section and, similarly, have no clear agreement with details of the experimental spectrum. This is not surprising, since commensurate stripes are long-range order (i.e., points in reciprocal space) best probed by a diffraction technique, whereas $\mu$SR, like NMR, is a point probe in real space. The histograms of Fig.~\ref{fig:homo-spct} are not as dense as those from stripe configurations, and there does not seem to be enough spectral weight at low frequencies, but the evidence for a striped AFM configuration from neutron diffraction~\cite{ZPBZ19u} is much more compelling.

\section{PARAMAGNETIC PHASE} \label{sec:paramag}

Two mechanisms for $\mu^+$ spin relaxation are expected in the high-temperature paramagnetic phase of a local-moment magnet: dynamic relaxation due to local-moment fluctuations, and (quasi)static relaxation from nuclear dipolar fields. In zero or low applied fields the resulting Kubo-Toyabe (KT) relaxation~\cite{KuTo67, *HUIN79} is due to the combined effects of these mechanisms. 

LF-$\mu$SR is a standard method for separation of the static and dynamic contributions~\cite{Brew94, Blun99, YaDdR11}, since a longitudinal field~$H_L$ ``decouples'' the $\mu^+$ spin polarization from the static local field distribution~\footnote{For large enough $H_L$ ($\gg$ the local-field distribution width) the resultant field is parallel to the $\mu^+$ spin polarization. Then there is no precession and thus no static relaxation, so that any observed relaxation is dynamic.}. Static $\mu^+$ relaxation functions~$G_s(H_L,t)$ for arbitrary $H_L$ (including zero field) have been determined for several local field distributions~\cite{KuTo67, UNYN81, *UYHS85, NoKa97, YaDdR11}. The combined effect of static and dynamic relaxation is often modeled by exponential damping of $G_s(H_L,t)$ with rate~$\lambda$:
\begin{equation} \label{eq:dampstat}
G(H_L,t) = e^{-\lambda t} G_s(H_L,t)\,.
\end{equation}

In La$_4$Ni$_3$O$_8$ $G(H_L,t)$ is complicated by the large number of inequivalent $\mu^+$ sites. This results in a broad inhomogeneous distribution of $\mu^+$ dipolar fields from Ni$^{1+}$ moments, as can be seen from the AFM-phase histograms discussed above. Similarly, the dipolar-field distribution from randomly-oriented $^{139}$La nuclei is further broadened because of the multiple $\mu^+$ sites. Obtaining $G(H_L,t)$ is then a difficult numerical problem even if these distributions were known, which is not the case for La$_4$Ni$_3$O$_8$ because of uncertainty in the $\mu^+$ sites.

We have carried out ZF- and LF-$\mu$SR experiments in the paramagnetic phase of La$_4$Ni$_3$O$_8$ between $T_N$ and 275~K. For analysis of the data we have chosen a rough approximate form of Eq.~(\ref{eq:dampstat}), using the static Lorentzian KT function~$G_s^\mathrm{Lor}(H_L,\alpha,t)$~\cite{UNYN81, UYHS85} with ``stretched exponential'' damping:
\begin{equation} \label{eq:selor}
G(H_L,\alpha,t) = \exp\left[-(\lambda^\ast t)^K\right]\, G_s^\mathrm{Lor}(H_L,\alpha,t) \,.
\end{equation}
We have taken a Lorentzian distribution of $^{139}$La dipolar fields with half-width~$\alpha/\gamma_\mu$ as an approximation beyond the Gaussian that is appropriate for a single $\mu^+$ site in a homogeneous system~\cite{HUIN79}. The LF static Lorentzian KT function is~\cite{UNYN81, UYHS85}
\begin{widetext}
\begin{eqnarray} \label{eq:LFlorKT}
G_s^\mathrm{Lor}%(H_L,\alpha,t) 
& = & 1 - \frac{\alpha}{\omega_L}j_1(\omega_L)\exp(-\alpha t) - \left(\frac{\alpha}{\omega_L}\right)^2\left[j_0(\omega_L t)\exp(-\alpha t) - 1\right] \nonumber\\ 
& & -\ \alpha\left[1+\left(\frac{\alpha}{\omega_L}\right)^2\right] \int_0^t j_0(\omega_L\tau)\exp(-\alpha\tau)\, d\tau\,,
\end{eqnarray}
\end{widetext}
where $\omega_L = \gamma_\mu H_L$, and $j_0$ and $j_1$ are spherical Bessel functions. In the ZF limit Eq.~(\ref{eq:LFlorKT}) simplifies to
\begin{equation} \label{eq:ZFlorKT}
G_s^\mathrm{Lor}(\alpha,t) = \frac{1}{3} + \frac{2}{3} (1 - \alpha t)\exp(-\alpha t)\,.
\end{equation}
The stretched-exponential form of the damping in Eq.~(\ref{eq:selor}) represents an inhomogeneous distribution of local dynamic relaxation rates~$\lambda_\mathrm{loc}$~\cite{KMCL96, John06}, where $\lambda^\ast$ is a characteristic rate (but not the average~\cite{John06}) and the stretching power~$K < 1$ characterizes the distribution function~$P(\lambda_\mathrm{loc})$ of the local rates. 

The stretched-exponential dynamic and Lorentzian static forms are crude approximations (both are strictly valid only for a $r^{-3}$ interaction with local moments in the dilute limit~\cite{TsHa68, UYHS85}). Their use is justified by their relative simplicity, and \textit{a~posteriori} by the fact that they give good fits to the data (see below). An alternative model for static relaxation in disordered systems, the ``Gaussian broadened Gaussian'' KT function~\cite{NoKa97}, does not fit well; for early times it varies as $t^2$ (i.e., as a Gaussian), whereas the data exhibit the linear early-time behavior of an exponential~\cite{UYHS85}.

\subsection{Zero field} \label{sec:exp-ZF}

\begin{figure}
\includegraphics[clip=,width=0.40\textwidth]{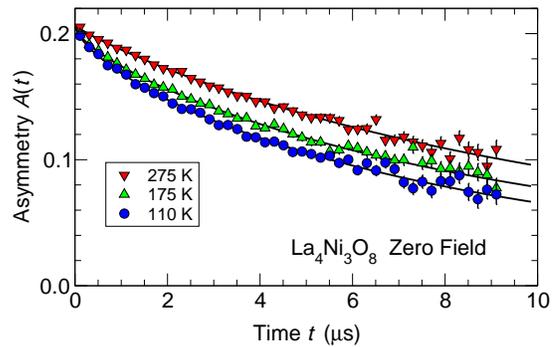}
\caption{\label{fig:exp-hitemp-asy}Representative zero-field $\mu$SR asymmetry relaxation in La$_4$Ni$_3$O$_8$, $T > T_N$. Legend: temperature values. Curves: fits of Eqs.~(\ref{eq:selor}) and (\ref{eq:ZFlorKT}) to the data.}
\end{figure}

\begin{figure}
\includegraphics[clip=,width=0.40\textwidth]{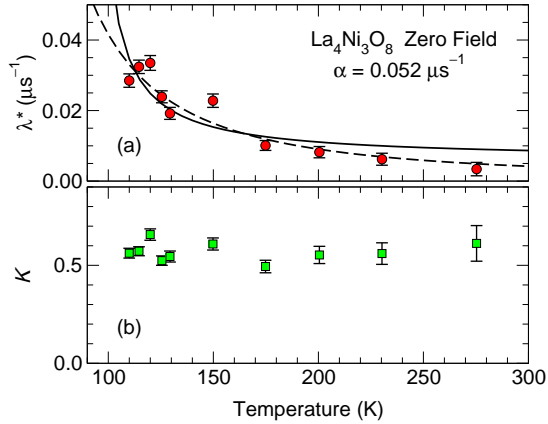}
\caption{\label{fig:res-ZF-relax} Temperature dependence of zero-field dynamic relaxation parameters in La$_4$Ni$_3$O$_8$, $T > T_N$. (a)~Stretched-exponential relaxation rate~$\lambda^\ast$. Solid curve: fit of $\lambda^\ast(T) \propto T/(T-T_0)$~\cite{MoUe74, MMP90} to the data, $T_0 = 90(1)$~K\@. Dashed curve: fit of Eq.~(\ref{eq:CO}) from the CO theory~\cite{ChOr90} to the data, $J = 86(22)$~K\@. (b)~Stretching power~$K$.}
\end{figure}

Representative ZF asymmetry spectra above $T_N$ are shown in Fig.~\ref{fig:exp-hitemp-asy}, and Fig.~\ref{fig:res-ZF-relax} shows the temperature dependence of $\lambda^\ast$ and $K$ in zero field from fits of Eqs.~(\ref{eq:selor}) and (\ref{eq:ZFlorKT}) to the data.
In the fits $\alpha$ was fixed at its 200-K value, but the results do not change qualitatively for $\pm$10\% variations of $\alpha$. The stretching power~$K(T)$ is essentially constant, whereas $\lambda^\ast(T)$ increases markedly with decreasing temperature as $T_N$ is approached.

The curves in Fig.~\ref{fig:res-ZF-relax}(a) give fits of theoretical expectations for 2D AFM spin fluctuations~\cite{MoUe74, MMP90, ChOr90}, as discussed in Ref.~\onlinecite{A-WDPG11} with respect to $^{139}$La NMR relaxation data. The solid curve is a phenomenological Curie-Weiss-like fit: $\lambda^\ast(T) \propto T/(T-T_0)$~\cite{MoUe74, MMP90}, which yields $T_0 = 90(1)$~K\@. The dashed curve in Fig.~\ref{fig:res-ZF-relax}(a) is the function
\begin{equation} \label{eq:CO}
\lambda^\ast(T) \propto x^{3/2}\,e^{1/x}/(1+x)^3\,,\quad x = T/1.13J \,,
\end{equation}
derived by Chakravarty and Orbach~\cite{ChOr90} (CO) from the theory of a 2D quantum Heisenberg antiferromagnet~\cite{CHN88, CHN89} mentioned in Sec.~\ref{sec:bkgnd}, in which the critical point is at $T = 0$. The weak inter-layer interaction that precipitates the 3D AFM transition in this picture is not expected to affect the critical behavior appreciably. Equation~(\ref{eq:CO}) is only valid for $T < 0.57J$, however~\cite{ChOr90}, where $J$ is the AFM exchange coupling. The fit value of~$J$ is 86(22)~K, so that temperatures in the paramagnetic phase of La$_4$Ni$_3$O$_8$ violate this condition. Thus the CO fit is inconsistent.

\subsection{\boldmath Longitudinal field, $T = 200$~K} \label{sec:200K} 

\begin{figure}
\includegraphics[clip=,width=0.40\textwidth]{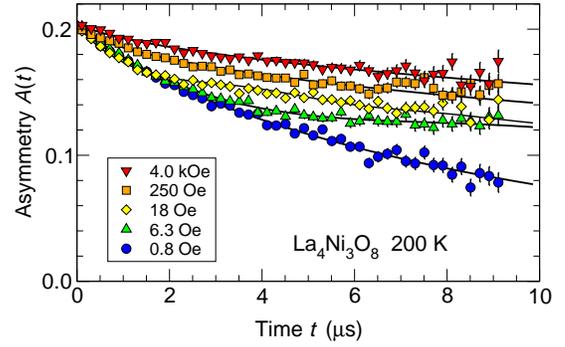}
\caption{\label{fig:exp-200K-asy}Representative longitudinal-field $\mu$SR asymmetry relaxation in La$_4$Ni$_3$O$_8$, $T = 200$~K\@. Legend: values of the longitudinal field~$H_L$. Curves: fits of Eqs.~(\ref{eq:selor})--(\ref{eq:LFlorKT}) to the data.}
\end{figure}

Figure~\ref{fig:exp-200K-asy} shows representative asymmetry spectra at 200~K for $0 \leqslant H_L \leqslant 4$~kOe. 
The curves are fits of Eqs.~(\ref{eq:selor})--(\ref{eq:LFlorKT}) to the data. The observed spectra are quite smooth, and the parameters~$\alpha$, $\lambda^\ast$, and $K$ in Eqs.~(\ref{eq:selor})--(\ref{eq:ZFlorKT}) are highly correlated statistically, making it hard to determine them separately. The following procedure was used: at high enough fields $G_s^\mathrm{Lor}(H_L,t) = 1$ independent of $\alpha$, so that stretched-exponential fits determine $\lambda^\ast$ and $K$. Assuming these are roughly field-independent, they were fixed and the lowest-field data were fit to determine $\alpha$. Then all three parameters were iteratively fixed and freed at low fields: $\alpha$ was fixed and $\lambda^\ast$ and $K$ freed, and vice versa, until convergence was obtained. Finally $\alpha$ was fixed, and $\lambda^\ast$ and $K$ were freed for fits at all fields. 

For low fields, the overall relaxation slows with increasing field but the early-time relaxation is not affected, which is a characteristic of static local-field decoupling. For higher fields the initial slope decreases, indicating a field dependence of $\lambda^\ast$ and/or $K$.

\begin{figure}
\includegraphics[clip=,width=0.40\textwidth]{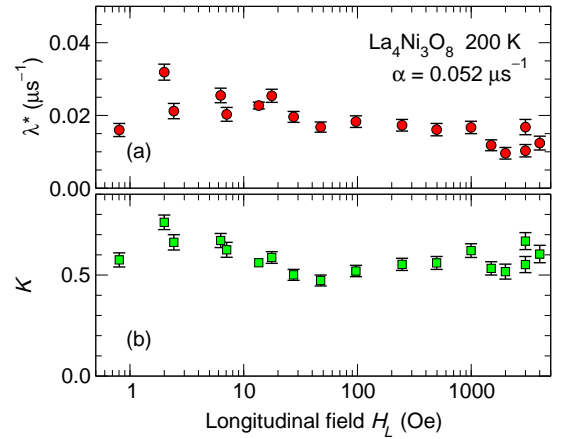}
\caption{\label{fig:res-200K-relax} Field dependence of Longitudinal-field dynamic relaxation parameters in La$_4$Ni$_3$O$_8$, $T = 200$~K\@. (a)~Stretched-exponential relaxation rate~$\lambda^\ast$. (b)~Stretching power~$K$.}
\end{figure}

The field dependences of $\lambda^\ast$ and $K$ at 200~K are shown in Fig.~\ref{fig:res-200K-relax}. 
The scatter in the parameters is obviously strongly correlated. The field dependences at low fields are probably artifacts of the assumed field independence of $\alpha$; a field dependence (as discussed in Ref.~\onlinecite{HUIN79} for the Gaussian relaxation rate at a single $\mu^+$ site) might be expected in La$_4$Ni$_3$O$_8$ because the $\mu^+$ frequency~$\gamma_\mu H_L$ is of the order of the lower of the two observed $^{139}$La quadrupolar splittings~\cite{PLNC10} for $H_L \sim 100$~Oe. Nevertheless, for $H_L \gtrsim 50$~Oe the fits no longer depend on $\alpha$, and the parameters are only weakly field-dependent. The stretching power~$K$ is close to the value~1/2 expected from a Lorentzian distribution of fluctuating fields in the motional-narrowing limit (rapid fluctuations)~\cite{TsHa68, UYHS85}, which is perhaps further justification for the use of Lorentzian distributions to describe $\mu^+$ relaxation in La$_4$Ni$_3$O$_8$.

\subsection{\boldmath Longitudinal field, $T =$ 110~K} \label{sec:exp-110K}

\begin{figure} [ht]
\includegraphics[clip=,width=0.40\textwidth]{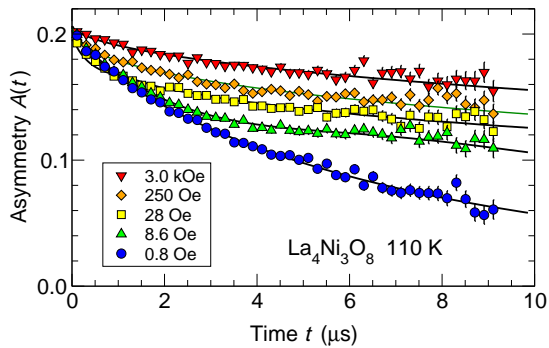}
\caption{\label{fig:exp-110K-asy}Representative longitudinal-field $\mu$SR asymmetry relaxation in La$_4$Ni$_3$O$_8$, $T = 110$~K\@. Legend: values of the longitudinal field~$H_L$. Curves: fits of Eqs.~(\ref{eq:selor})--(\ref{eq:LFlorKT}) to the data.} 
\end{figure}

\begin{figure} [ht]
\includegraphics[clip=,width=0.40\textwidth]{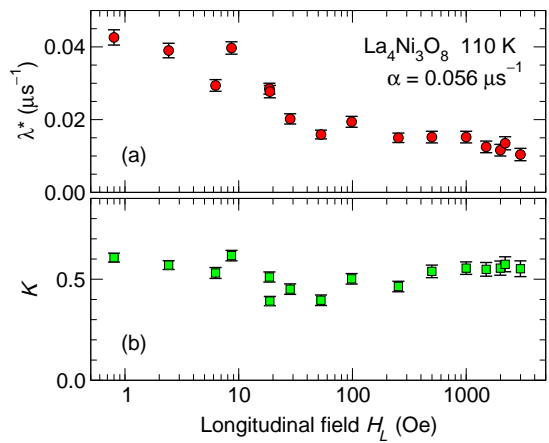}
\caption{\label{fig:res-110K-relax} Field dependence of Longitudinal-field dynamic relaxation parameters in La$_4$Ni$_3$O$_8$, $T = 110$~K\@. (a)~Stretched-exponential relaxation rate~$\lambda^\ast$. (b)~Stretching power~$K$.}
\end{figure}

A similar analysis was applied to LF data taken for $T = 110$~K, a temperature just above $T_N$\@. Representative asymmetry spectra are shown in Fig.~\ref{fig:exp-110K-asy}, and the field dependences of $\lambda^\ast$ and $K$, determined as described above for 200~K, are shown in Fig.~\ref{fig:res-110K-relax}. 
The values of $K$ at the two temperatures are comparable, but at 110~K a clear increase of $\lambda^\ast$ at low fields is observed. This increase and the temperature independence of $\lambda^\ast$ at high fields are discussed further in Sec.~\ref{sec:slowing}.

Values of $\lambda^\ast$ for $H_L = 0.8$~Oe at $T = 200$~K and 110~K (Figs.~\ref{fig:res-200K-relax} and \ref{fig:res-110K-relax}) are larger than the ZF values at those temperatures (Fig.~\ref{fig:res-ZF-relax}). This field dependence at low fields is not presently understood.

\section{DISCUSSION} \label{sec:disc}

\subsection{AFM transition and spin configuration}

The complex zero-field $\mu$SR spectra observed below 105~K in La$_4$Ni$_3$O$_8$ (Fig.~\ref{fig:res-FFT}) clearly indicate the magnetic nature of the transition. The multiple frequencies arise from a number of magnetically inequivalent $\mu^+$ stopping sites in the crystal. The sharp peaks indicate commensurate magnetic order, since in an incommensurate structure crystallographically equivalent $\mu^+$ stopping sites would be subject to a continuous distribution of frequencies. The smooth differences in temperature dependencies of the various peak frequencies between $T_M$ and $\sim$25~K [Fig.~\ref{fig:res-6freq}(a)] suggest reorientation of the AFM magnetic structure with decreasing temperature below $T_N$.

\subsubsection{Abrupt onset at $T_N$} 

From Figs.~\ref{fig:res-FFT}--\ref{fig:res-6freq} we conclude that the transition at $T_N = 105$~K has an abrupt onset, since the peak frequencies appear discontinuously at $T_N$\@. This is suggestive of a first-order transition. It is, however, also consistent with a continuous second-order transition in an inhomogeneous system~\cite{MVBdR94}, if the intrinsic onset of the order parameter is sufficiently abrupt (i.e., the critical exponent~$\beta$ is sufficiently small) compared to the width of the inhomogeneous transition temperature distribution. A simple model for zero-field $\mu$SR (and NQR) spectra in this situation, described in Appendix~\ref{app:model}, shows that near $T_N$ the spectra become broadly distributed, with reduced-amplitude peaks at their maxima. Then the transition appears discontinuous (Fig.~\ref{fig:spectra}), as we observe in La$_4$Ni$_3$O$_8$ (Fig.~\ref{fig:res-FFT}). We cannot discriminate between this and a true first-order discontinuity on the basis of our data (the term ``abrupt'' describes both possibilities).

\paragraph{Comparison with neutron diffraction.\label{par:neucomp}} A commensurate AFM phase is found in La$_4$Ni$_3$O$_8$ by neutron Bragg diffraction~\cite{ZPBZ19u}, but with a smooth continuous transition. This discrepancy with the $\mu$SR result can be understood in the above scenario~\cite{IJH93, MVBdR94}. The effect of the distribution is quite different in the two techniques, due to the fact that magnetic resonance experiments probe real space whereas neutron diffraction probes reciprocal space. In neutron diffraction the AFM Bragg intensity is proportional to the volume fraction of magnetic order, and therefore increases smoothly as the temperature is decreased through an inhomogeneous distribution of transition temperatures.

\paragraph{Comparison with NMR.} The $\mu$SR evidence for an abrupt transition in La$_4$Ni$_3$O$_8$ is also in conflict with the smooth transition observed in $^{139}$La NMR spectra [Ref.~\onlinecite{PLNC10} (suppl.)]. The NMR spectra were obtained from a partially-oriented powder sample in an applied field of 42 kOe~rather than zero field, and the conclusion of a smooth transition is mainly based on a single point at $\sim$95~K\@. It seems possible that partial random orientation and/or the applied field affected the results in these experiments.

\paragraph{Comparison with other materials.} $\mu$SR studies of the doped nickelate~La$_{2-x}$Sr$_x$NiO$_4$~\cite{CBPH96, JCBH99, Klau04} yield an abrupt but continuous frequency onset at $T_N$ with a small value of the critical exponent~$\beta$, similar to our results in La$_4$Ni$_3$O$_8$.

Discontinuous transitions were observed in the layered AFM cuprate~La$_2$CuO$_{4+\delta}$ in $^{139}$La NQR~\cite{BDIS90} and other hyperfine experiments~\cite{UKYK87b, TXST90}, but not in neutron diffraction; the discrepancy can be understood as noted above in Sec.~\ref{par:neucomp}. Subsequent $^{139}$La NQR experiments in a sufficiently homogeneous La$_2$CuO$_{4+\delta}$ crystal~\cite{MVBdR94} resolved an ``abrupt but continuous'' change of frequency at the transition. Our results in La$_4$Ni$_3$O$_8$ might also be due to such an inhomogeneity-induced discontinuity, and there is a clear need for experiments on more homogeneous samples of this compound.

\subsubsection{AFM spin configuration} \label{sec:spinconf}

Histograms of calculated Ni-ion dipolar fields at candidate $\mu^+$ sites for a number of charge stripe models and AFM configurations (Sec.~\ref{sec:modelcalc}, Appendix~\ref{app:stripes}) capture the general features of frequency spectra and suggest reduced Ni moments (0.4--0.5$\mu_B$), but do not uniquely determine the specific AFM configuration. Dipolar-field histograms for AFM configurations without stripes have somewhat less structure than the observed spectra. 

The broadening of Peak~1 below $\sim$25~K [Fig.~\ref{fig:res-6freq}(d)] is evidence for the onset of significant disorder at low temperatures. The low-frequency peaks arise from $\mu^+$ sites far from Ni trilayers, suggesting that the disorder is associated with the lack of $c$-axis correlation reported in Ref.~\cite{ZPBZ19u}. The absence of broadening for Peak~6, which is from $\mu^+$ sites in or near the Ni trilayers, shows that the AFM state there is commensurate and not strongly disordered at low temperatures. Peak~1 becomes narrow above 25~K (Fig.~\ref{fig:res-FFT}), which indicates motional averaging of the corresponding local field. 

This behavior suggests that the 105-K transition is within the trilayers and not between them. We speculate that at intermediate temperatures, each trilayer fluctuates dynamically as a whole but independently of neighboring trilayers, maintaining nearby $\mu^+$ local fields more or less constant. Such ``stripe fluctuation'' has been reported in La$_{2-x}$Sr$_x$NiO$_4$~\cite{Klau04}. Below 25~K a weaker (possibly anisotropic) inter-trilayer interaction leads to disordered full 3D spin freezing, with a broad distribution of local fields at $\mu^+$ sites distant from trilayers.

\subsection{Paramagnetic-phase relaxation; critical slowing} \label{sec:slowing}

As discussed in Sec.~\ref{sec:exp-ZF}, increase of the ZF $\mu^+$ relaxation rate~$\lambda^\ast(T)$ as $T \to T_N$ from the paramagnetic phase [Fig.~\ref{fig:res-ZF-relax}(a)] suggests critical slowing of spin fluctuations associated with the transition. The data can be fit with Eq.~(\ref{eq:CO}) from the CO theory~\cite{ChOr90}, in which the critical fluctuations are 2D in nature and diverge as $T \to 0$. The value of the exchange constant~$J$ from this fit is lower than the temperatures of measurement, however, which invalidates the fit. A Curie-Weiss-like fit yields a divergence at 90(1)~K, suggesting that it is associated with $T_N$ rather than zero temperature.

Curie-Weiss and CO-theory fits [Fig.~\ref{fig:res-ZF-relax}(a)] both yield characteristic temperatures~($T_0$ and $J$, respectively) of the order of $T_N$. This seems compatible with a connection between critical slowing and the 3D phase transition, although, as noted above, the CO fit is not valid.

Comparison of the field dependencies of $\lambda^\ast(T)$ at 200~K [Fig.~\ref{fig:res-200K-relax}(a)] and 110~K [Fig.~\ref{fig:res-110K-relax}(a)] suggests that the critical point is at zero field. The high-field data at both temperatures have similar magnitudes, but the low-field relaxation increases strongly with decreasing field at the lower temperature (which is close to $T_N$). 

\paragraph{Comparison with NMR.} Like the $\mu$SR data, the paramagnetic-phase $^{139}$La nuclear spin-lattice relaxation rate in La$_4$Ni$_3$O$_8$~\cite{A-WDPG11} can be fit with a Curie-Weiss law, but with a divergence as $T \to 0$ that disagrees strongly with the $\mu$SR result. The NMR data were also fit by Eq.~(\ref{eq:CO}), which was taken as evidence for the CO theory as discussed above. However, the exchange constant 129(5)~K from the NMR data is, like that from the $\mu$SR data, of the order of $T_N$, and therefore the temperatures of measurement ($T > T_N$) do not obey the condition~$T < 0.69J$ required for self-consistency of the theory~\cite{ChOr90}. Furthermore, the NMR experiments were carried out in a field of 42~kOe on very broad lines, which make the measurement difficult. For nonzero but much lower fields (0.1--4~kOe, Figs.~\ref{fig:res-200K-relax} and \ref{fig:res-110K-relax}) $\mu^+$ spin relaxation rates show no sign of critical slowing between 200~K and 110~K as noted above. 

\paragraph{Specific heat peak.} A peak in the temperature dependence of the specific heat is observed at $T_N$~\cite{PLNC10, ZCPZ16}. We note that that a specific-heat ``peak'' could also be due to first-order latent heat in an inhomogeneous sample with a spread of transition temperatures. Furthermore, charge ordering in La$_4$Ni$_3$O$_8$ might contribute to the specific heat ($\mu$SR is not directly sensitive to charge ordering). Nevertheless, the peak has been taken as evidence for a second-order transition~\cite{A-WDPG11}, which is not in conflict with the $\mu$SR relaxation data.

\paragraph{Comparison with other materials.} Critical slowing with a $T \to 0$ divergence was observed in NQR experiments on La$_2$CuO$_{4+\delta}$~\cite{ISYK93}, and taken as evidence for the CO theory~\cite{ChOr90}. Fits of Eq.~(\ref{eq:CO}) to the NQR data yield $T \ll J$~\cite{ISYK93}, consistent with applicability of the theory. 

La$_4$Ni$_3$O$_8$ is intrinsically overdoped, and a more appropriate comparison might be with results from doped La$_2$CuO$_4$ with charge ordering. Previous $^{63,65}$Cu and $^{139}$La NQR studies of underdoped La$_{2-x}$Sr$_x$CuO$_4$, $0 \leqslant x \leqslant 0.15$,~\cite{ISYK93} found that critical slowing for $x = 0$ at high temperatures is rapidly suppressed by Sr doping. Recent NMR experiments~\cite{ITAA17, ATIH17} using a single crystal of La$_{1.885}$Sr$_{0.115}$CuO$_4$ revealed inhomogeneity and considerable structure associated with staging and oxygen diffusion at low temperatures, but did not revise the earlier conclusions. A NMR study~\cite{ImLe18} of single-crystal La$_2$CuO$_{4+y}$, $y \sim 0.11$, showed Curie-Weiss behavior of the $^{63}$Cu relaxation rate for temperatures $\gtrsim 60$~K\@. This result is similar to both NMR and ZF-$\mu$SR data from La$_4$Ni$_3$O$_8$, but the doping level is higher in the nickelate. To our knowledge no $\mu$SR studies of critical slowing in the paramagnetic phases of La$_2$CuO$_4$-based systems, doped or undoped, have been reported.

\section{CONCLUSIONS} \label{sec:concl}

Comparison of $\mu$SR spectra in the AFM phase of La$_4$Ni$_3$O$_8$ with calculations of dipolar fields at candidate $\mu^+$ sites does not determine the AFM configuration uniquely, but is consistent with the stripe structure from neutron diffraction and suggests a reduced ordered Ni$^{1+}$ moment~0.4--$0.5\mu_B$.

The nature of the 105-K transition is not well determined. Critical slowing is not expected for a first-order transition, however, and its observation is probably the best evidence that the transition in La$_4$Ni$_3$O$_8$ is ``abrupt but continuous''~\cite{MVBdR94}. 

The behavior of Peaks~1 and 6 in the AFM spectra (Figs.~\ref{fig:res-FFT} and \ref{fig:res-6freq}) is strong evidence for commensurate AFM trilayers that fluctuate dynamically at temperatures between 25~K and $T_N$, and freeze with uncorrelated $c$-axis disorder below $\sim$25 K.

There is no evidence for 2D critical slowing with a divergence as $T \to 0$ in $\mu$SR data from La$_4$Ni$_3$O$_8$. The low-frequency spin dynamics of La$_4$Ni$_3$O$_8$ and La$_2$CuO$_{4+\delta}$ are not comparable, in spite of some similarities. 

More work, both theoretical and experimental, is clearly necessary to understand the phase diagram of this enigmatic material.

\begin{acknowledgments}
We are grateful for the assistance of B. Hitti and D. Arsenau of the TRIUMF Centre for Molecular and Materials Research during these experiments. One of us (D.E.M.) thanks J.~Spalding for useful comments. This research was supported in part by the National Science Foundation under the Cal State LA/Penn State PREM Program, NSF Grants~DMR-1523588% (O.O.B.)
, DMR-1809306% (V.V.P.)
, and DMR-1506677% (P.-C.H.)
, by the National Natural Science Foundation of China No.~11774061% (L.S.)
, and by the University of California, Riverside, Academic Senate% (D.E.M.)
.
\end{acknowledgments}

 \raggedbottom
\appendix 

\section{Other AFM stripe models} \label{app:stripes}

\begin{figure} [ht]
\includegraphics[clip=,width=0.35\textwidth]{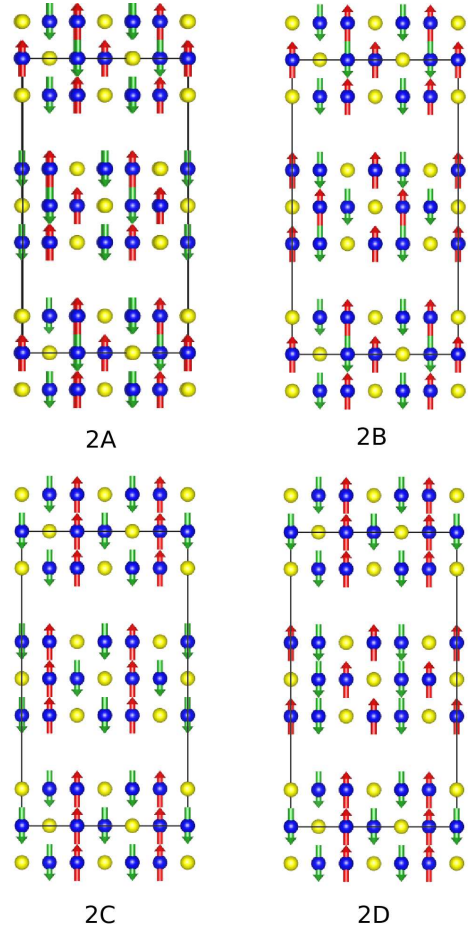}
\caption{\label{fig:AFM-mod2}AFM moment Configurations A--D in La$_4$Ni$_3$O$_8$ on stripe Model~2~\cite{ZCPZ16}. See Fig.~\ref{fig:AFM-mod1} for designations.}
\end{figure}

\begin{figure} %[ht]
\includegraphics[clip=,width=0.35\textwidth]{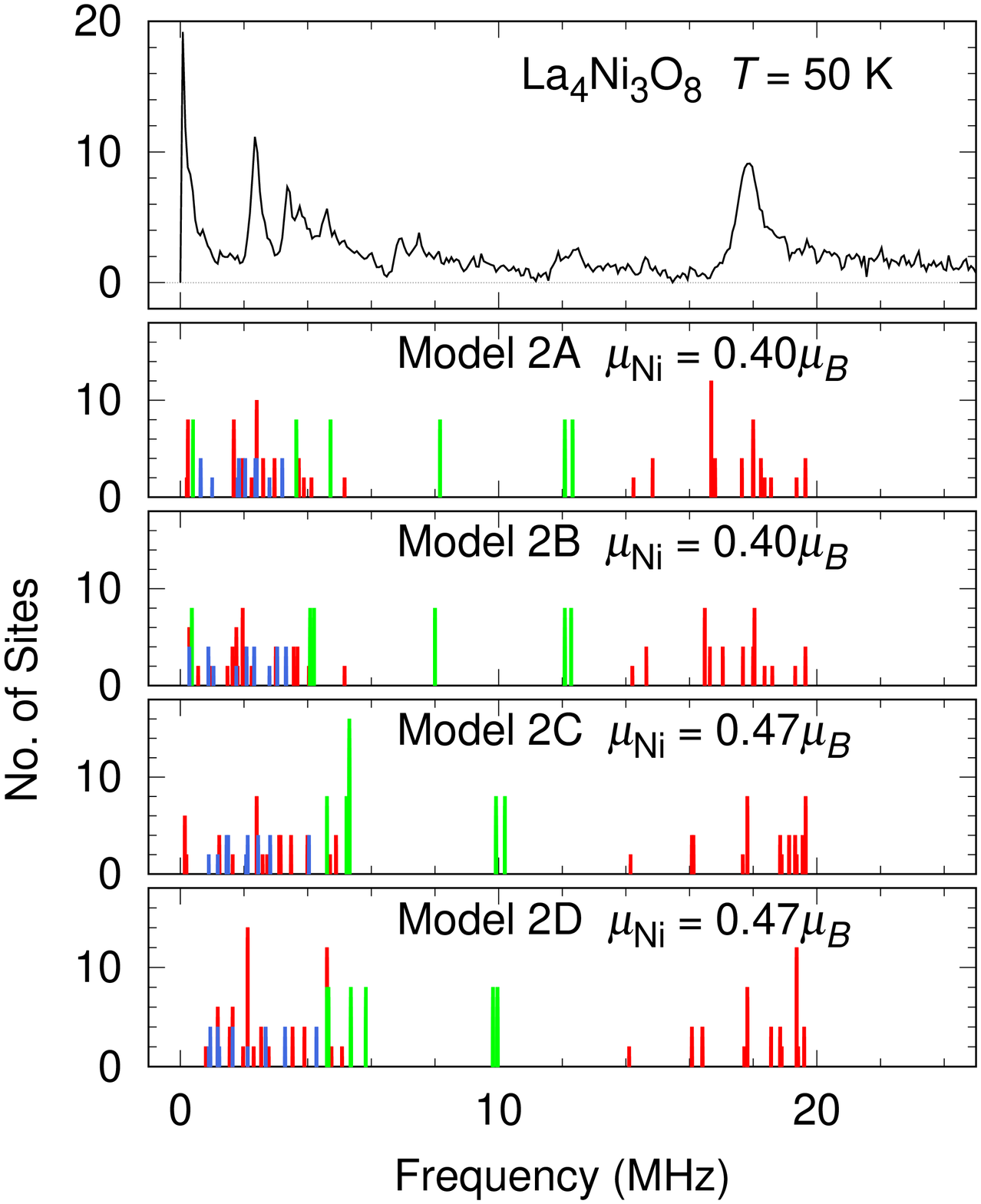}
\caption{\label{fig:disc-spct2} Histograms of dipolar-field $\mu^+$ frequencies from AFM stripe Models~2A--2D compared to 50-K $\mu$SR FT spectrum in La$_4$Ni$_3$O$_8$. Colors as in Fig.~\ref{fig:disc-spct1}.}
\end{figure}

\begin{figure} %[ht]
\includegraphics[clip=,width=0.35\textwidth]{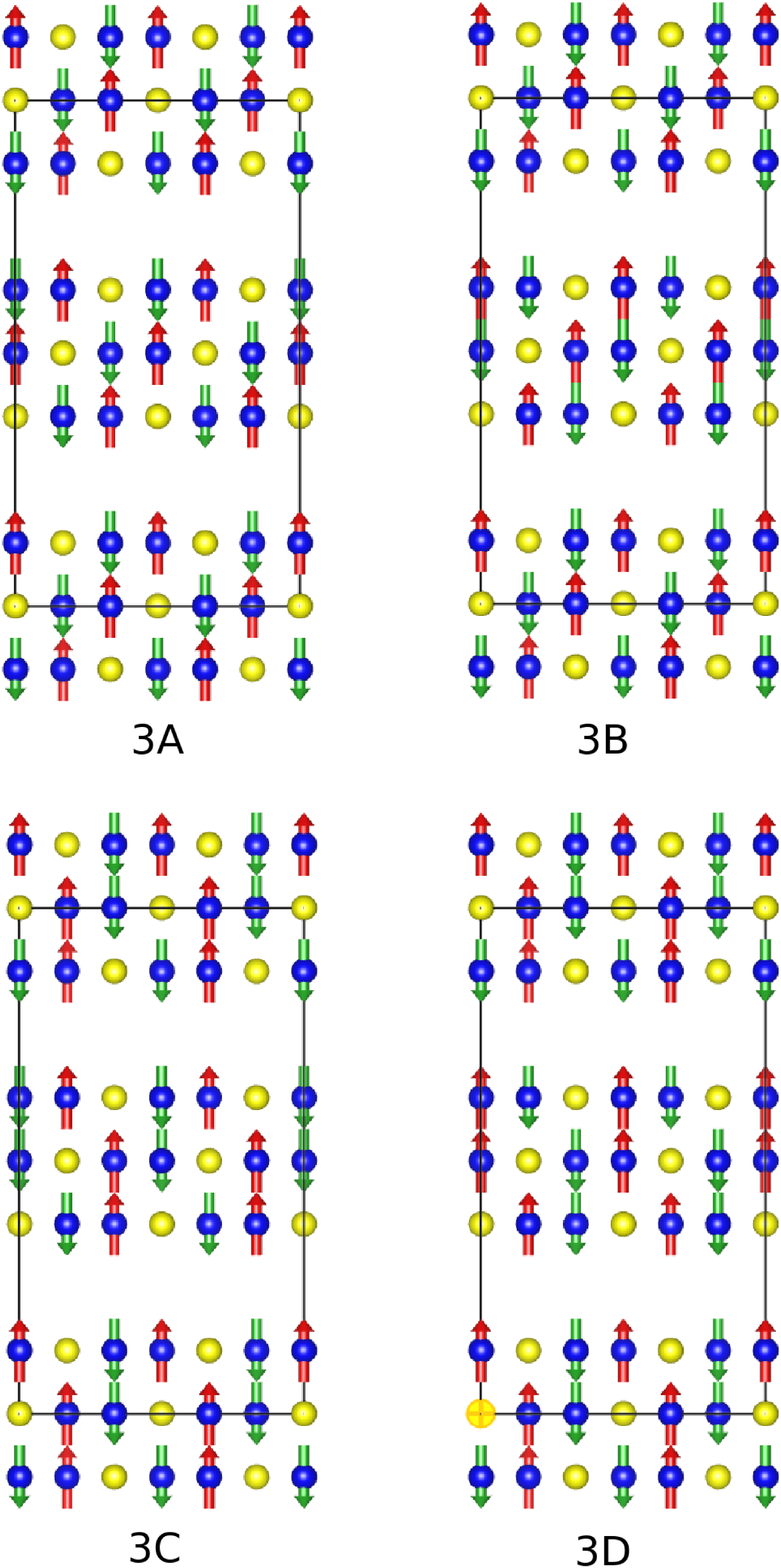}
\caption{\label{fig:AFM-mod3}AFM moment Configurations A--D in La$_4$Ni$_3$O$_8$ on stripe Model~3~\cite{ZCPZ16}. See Fig.~\ref{fig:AFM-mod1} for designations.}
\end{figure}

\begin{figure} %[ht]
\includegraphics[clip=,width=0.35\textwidth]{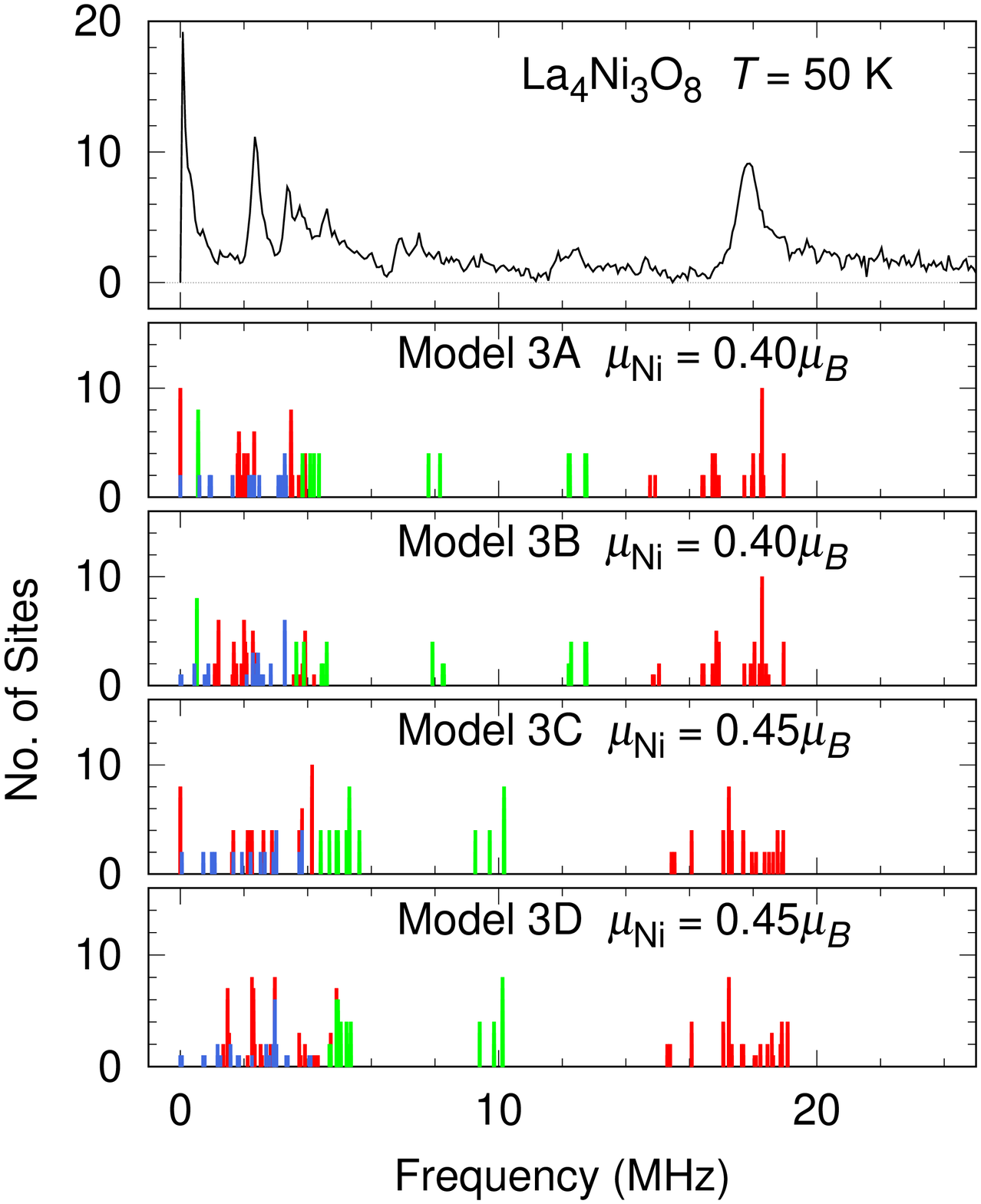}
\caption{\label{fig:disc-spct3} Histograms of dipolar-field $\mu^+$ frequencies from AFM stripe Models~3A--3D compared to 50-K $\mu$SR FT spectrum in La$_4$Ni$_3$O$_8$. Colors as in Fig.~\ref{fig:disc-spct1}.}
\end{figure}

\begin{figure} %[ht]
\includegraphics[clip=,width=0.35\textwidth]{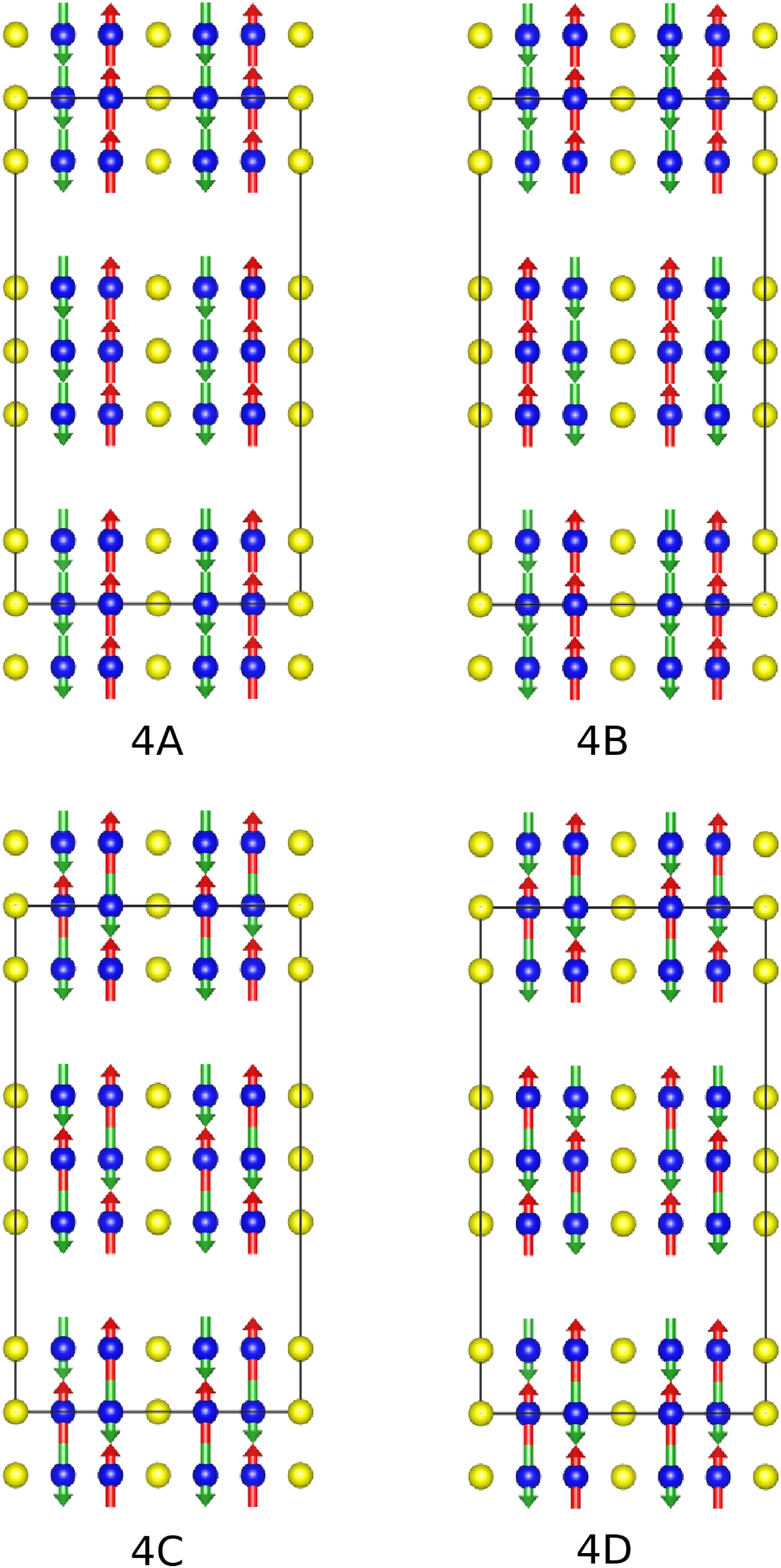}
\caption{\label{fig:AFM-mod4}AFM moment Configurations A--D in La$_4$Ni$_3$O$_8$ on stripe Model~4~\cite{ZCPZ16}. See Fig.~\ref{fig:AFM-mod1} for designations.}
\end{figure}

\begin{figure} %[ht]
\includegraphics[clip=,width=0.35\textwidth]{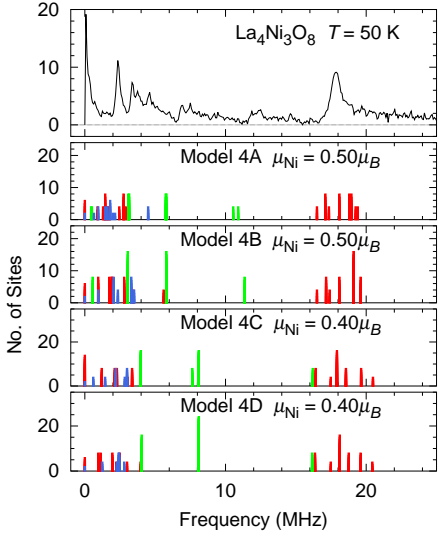}
\caption{\label{fig:disc-spct4} Histograms of dipolar-field $\mu^+$ frequencies from AFM stripe Models~4A--4D compared to 50-K $\mu$SR FT spectrum in La$_4$Ni$_3$O$_8$. Colors as in Fig.~\ref{fig:disc-spct1}.}
\end{figure}

AFM configurations %defined in Sec.~\ref{sec:modelcalcstripes} 
and corresponding dipolar-field histograms for stripe Models~2--4 are shown in Figs.~\ref{fig:AFM-mod2}--\ref{fig:disc-spct4}. The trends seen in Model~-1 histograms and discussed in Sec.~\ref{sec:modelcalcstripes} (low Ni moments, low- and high-frequency groups, only small changes for inverted alternate trilayers) continue to be found.

\pagebreak[4] 
\section{Inhomogeneity and the spontaneous frequency distribution for abrupt transitions} \label{app:model}

We consider a simple model for a system with an abrupt second-order magnetic transition, i.e., a small value of the order-parameter critical exponent~$\beta$. A first-order transition can be simulated by letting $\beta \to 0$. 

The temperature dependence of a spontaneous zero-field frequency~$f$ (proportional to the order parameter) is assumed to be of the critical-exponent form
\begin{equation} \label{eq:fofT}
f(T,T_c) = f_0(1 - T/T_c)^\beta,\ 0 < T < T_c \,,
\end{equation}
where $f_0$ is the zero-temperature value. At a temperature~$T$ the frequency distribution function~$P(f,T)$ (the spectrum) is given by
\begin{equation}
P(f,T) = \left\{\begin{array}{l}
 {\displaystyle P[T_c(f,T)]\,\frac{dT_c(f,T)}{df}} \,,\ f_\mathrm{min} < f < f_\mathrm{max} \,, \\
 0\ \text{otherwise} \,,
 \end{array}
\right.
\end{equation}
where from Eq.~(\ref{eq:fofT}) 
\begin{equation}
T_c(f,T) = \frac{T}{1 - (f/f_0)^{1/\beta}} \,,
\end{equation}
and $P(T_c)$ is the inhomogeneous distribution of transition temperatures~$T_c$, taken to be nonzero for $T_c^\mathrm{min} < T_c < T_c^\mathrm{max}$. At a given temperature~$T$, $P(f,T)$ lies between the limits~$f_\mathrm{min\,(max)}(T) = f(T,T_c^{\mathrm{min\,(max)}})$. This yields
\begin{equation} \label{eq:Poff}
P(f,T) = \left\{\begin{array}{l}
 {\displaystyle \frac{P[T_c(f,T)]\,T(f/f_0)^{1/\beta}}{\beta f[1 - (f/f_0)^{1/\beta}]^2}} \,,\ f_\mathrm{min} < f < f_\mathrm{max} \,, \\
 0\ \text{otherwise} \,.
 \end{array}
\right.
\end{equation}

In our simple model $P(T_c$) is taken to be uniform with mean~$T_{c0}$ and half width~$\Delta T_c$:
\begin{equation}
P(T_c) = \left\{\begin{array}{l}
 {\displaystyle \frac{1}{2\Delta T_c}}\,,\ T_c^\mathrm{min} < T_c < T_c^\mathrm{max}\,, \\
 0\ \text{otherwise} \,,
 \end{array}
\right.
\end{equation}
 with $T_c^\mathrm{min\,(max)} = T_{c0} \mp \Delta T_c$.  
\begin{figure} [H] \centering
\includegraphics[clip=,width=0.40\textwidth]{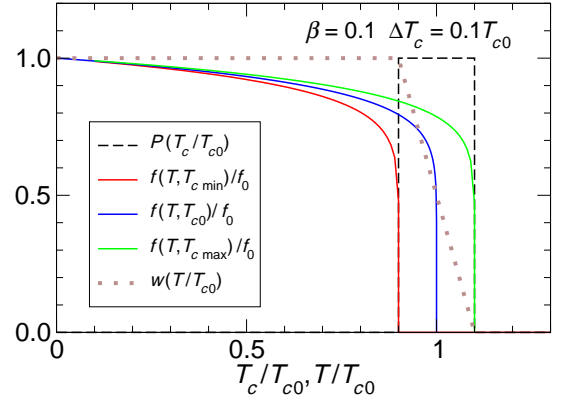}
\caption{\label{fig:Tc+dist} Dashed curve: temperature dependence of transition temperature distribution~$P(T_c$), normalized by $2\Delta T_c$ for display. Solid curves: temperature dependencies of spontaneous NMR/NQR/$\mu$SR frequencies~$f(T,T_c)$ for $T_c^\mathrm{min}$ (red), $T_{c0}$ (blue), and $T_c^\mathrm{max}$ (green). Dotted curve: spectral weight~$w(T)$.}
\end{figure}
Figure~\ref{fig:Tc+dist} shows $P(T_c)$, $f(T,T_c)$ for $T_c = \ T_c^\mathrm{min} \,,\ T_{c0} \,,$ and $T_c^\mathrm{max}$, and the normalized spectral weight
\begin{equation}
w(T) = \left\{\begin{array}{l}
 1 - {\displaystyle \int_{T_c^\mathrm{min}}^T P(T_c)\,dT_c}\,,\ 
 T_c^\mathrm{min} < T < T_c^\mathrm{max}\,, \\
 1\ \text{otherwise} \,.
 \end{array}
\right.
\end{equation}
\begin{figure} [H] \centering
\includegraphics[clip=,width=0.40\textwidth]{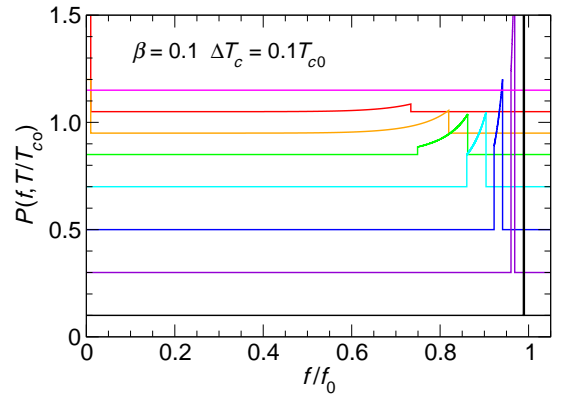}
\caption{\label{fig:spectra} Representative zero-field frequency spectra from a uniform distribution of second-order magnetic phase transition temperatures. Spectra are normalized by 0.01 for display, and are offset so that each baseline is at the corresponding temperature~$T/T_{c0}$ on the $P$ axis. For $T > T_c^\mathrm{min}$ spectral weight is transferred from nonzero frequencies to $f = 0$.}
\end{figure}
The integral is the fraction~$1 - w(T)$ of the sample volume with $T_c < T$, for which $f = 0$. 

Representative spectra from Eq.~(\ref{eq:Poff}) are shown in Fig.~\ref{fig:spectra}. Their behavior is qualitatively similar to that of Peak~6 of our experimental results in La$_4$Ni$_3$O$_8$ (Fig.~\ref{fig:res-FFT}): common features include a slight frequency decrease and loss of $f > 0$ spectral weight (transferred to $f = 0$) for $T > T_c^\mathrm{min}$. The remaining weight for $f > 0$ becomes broadly distributed, with a peak at the maximum. The shapes of the observed and calculated spectra differ, but these depend strongly on the form of the $T_c$ distribution and would not be expected to be reproduced in this simple model. The observed peaks are broader than those of the model, probably reflecting disorder that is not taken into account in the model.

%\bibliography{La4Ni3O8}

%apsrev4-2.bst 2019-01-14 (MD) hand-edited version of apsrev4-1.bst
%Control: key (0)
%Control: author (8) initials jnrlst
%Control: editor formatted (1) identically to author
%Control: production of article title (0) allowed
%Control: page (0) single
%Control: year (1) truncated
%Control: production of eprint (0) enabled
%

\end{document}